\documentclass{amsart}
\usepackage{latexsym}
\usepackage{amssymb}
\usepackage{mathrsfs}
\theoremstyle{plain}
\newtheorem{thm}{Theorem}
\newtheorem{prop}{Proposition}
\newtheorem{lemma}{Lemma}

\newtheorem{cor}{Corollary}
\theoremstyle{definition}
\newtheorem{definition}{Definition}

\newcommand{\Spp}{\ensuremath{\Sigma_{+}}}
\newcommand{\Sm}{\ensuremath{\Sigma_{-}}}
\newcommand{\No}{\ensuremath{N_{1}}}
\newcommand{\Nt}{\ensuremath{N_{2}}}
\newcommand{\Nth}{\ensuremath{N_{3}}}

\renewcommand{\a}{\ensuremath{\alpha}}
\newcommand{\e}{\ensuremath{\epsilon}}
\newcommand{\tx}{\ensuremath{\tilde{x}}}
\renewcommand{\th}{\ensuremath{\tilde{h}}}
\newcommand{\ty}{\ensuremath{\tilde{y}}}
\newcommand{\tbx}{\ensuremath{\tilde{\mathbf{x}}}}

\newcommand{\tg}{\ensuremath{\tilde{g}}}
\newcommand{\tsl}{\tilde{\mathrm{Sl}}(2,\mathbb{R})}
\newcommand{\psl}{\mathrm{PSl}(2,\mathbb{R})}
\renewcommand{\sl}{\mathrm{Sl}(2,\mathbb{R})}

\renewcommand{\th}{\tilde{h}}
\newcommand{\tga}{\tilde{\gamma}}

\begin{document}
\title{Future asymptotic expansions of Bianchi VIII vacuum metrics}
\author{Hans Ringstr\"{o}m}
\address{Max-Planck-Institut f\"{u}r Gravitationsphysik, Am M\"{u}hlenberg 1,
D-14476 Golm, Germany}

\begin{abstract}
Bianchi VIII vacuum solutions to Einstein's equations are causally
geodesically complete to the future, given an appropriate
time orientation, and the objective of this article is to analyze
the asymptotic behaviour of solutions in this time direction.
For the Bianchi class A spacetimes, there is a formulation
of the field equations that was presented in an article
by Wainwright and Hsu, and in a previous article we analyzed
the asymptotic behaviour of solutions in these variables. 
One objective of this paper is to give an asymptotic expansion
for the metric. Furthermore, we relate this expansion to the
topology of the compactified spatial hypersurfaces of homogeneity.
The compactified spatial hypersurfaces have the topology of Seifert fibred
spaces and we prove that in the case of NUT Bianchi VIII spacetimes,
the length of a circle fibre converges to a positive constant
but that in the case of general Bianchi VIII solutions, the length
tends to infinity at a rate we determine. Finally, we give
asymptotic expansions for general Bianchi VII${}_{0}$ metrics.
\end{abstract}
\maketitle

\section{Introduction}

In a previous article \cite{jag3}, we considered the Bianchi class A
vacuum cosmologies in the expanding direction. In it, we analyzed the 
asymptotics in terms of the variables of Wainwright and Hsu (note 
also that an analysis of Bianchi VIII in the case where matter is
present has been carried out in \cite{wain}). One
may then ask what purpose the present article serves. The point is
that the spacetime metric can be written
\begin{equation}\label{eq:metric}
\bar{g}=-d t^{2}+\sum_{i=1}^{3}a_{i}^{2}(t)\xi^{i}\otimes \xi^{i}
\end{equation}
on $I\times G$, where $G$ is a three dimensional Lie group and the 
$\xi^{i}$ are dual to a left invariant basis of the tangent bundle of 
$G$. The $a_{i}$ can be computed using the variables of Wainwright and 
Hsu, but the computation involves an integration.
In particular, in the case of Bianchi VII${}_{0}$ and VIII, the
relevant integrand
tends to zero for some of the $a_{i}$, and thus a further analysis is
necessary. A second question of interest which was left open in
\cite{jag3} is the question of how the geometry of the spatial
hypersurfaces of homogeneity behaves with time, and how this connects
with topology. Recently, general conjectures on how the future asymptotics 
of solutions to Einstein's equations should be have emerged, see
e.g. \cite{fimon} and \cite{anderson}. We do not wish to describe
these ideas in general here, but would like to state their
implications for Bianchi VIII. The compactified spatial hypersurfaces 
have the structure of Seifert fibred spaces, and the conjecture is
that by rescaling the induced metric on the spatial hypersurfaces
by dividing by proper time squared for example, the circle fibres in
the Seifert fibration should collapse. Here we prove more than this,
in fact we are able to give a rate of collapse and to distinguish
between general Bianchi VIII spacetimes and the subclass of NUT
spacetimes by considering how the length of a circle fibre evolves
with time. Finally, note that the results presented here are of
interest when trying to redo the analysis in \cite{chom} for
non-trivial circle bundles over a higher genus surface.

In order to explain the results it is necessary to introduce some 
terminology.  Let $G$ be a $3$-dimensional Lie
group and $e_{i}$, $i=1,2,3$ be a basis of the Lie algebra with structure
constants determined by  $[e_{i},e_{j}]=\gamma_{ij}^{k}e_{k}$. If
$\gamma_{ik}^{k}=0$, then the Lie algebra and Lie group are said to 
be of \textit{class A} and  
\begin{equation}\label{eq:sconstants}
\gamma_{ij}^{k}=\epsilon_{ijm}n^{km}
\end{equation}
where the symmetric matrix $n^{ij}$ is given by
\begin{equation}\label{eq:ndef}
n^{ij}=\frac{1}{2}\gamma^{(i}_{kl}\epsilon_{}^{j)kl}.
\end{equation}
\begin{definition}\label{def:data}
\textit{Class A vacuum initial data} for Einstein's equations consist of
the following. A  Lie group $G$ of class A, a left invariant metric
$g$ on $G$ and a left invariant symmetric covariant two-tensor $k$ on
$G$ satisfying
\begin{equation}\label{eq:con1}
R_{g}-k_{ij}k^{ij}+(\mathrm{tr}_{g} k)^2=0
\end{equation}
and
\begin{equation}\label{eq:con2}
\nabla_{i}\mathrm{tr}_{g} k-\nabla^{j}k_{ij}=0
\end{equation}
where $\nabla$ is the Levi-Civita connection of $g$ and $R_{g}$ is
the corresponding scalar curvature, indices are raised and lowered
by $g$.
\end{definition}

\begin{definition}\label{def:can}
Consider class A vacuum initial data $(G,g,k)$. A left invariant
basis of the tangent bundle $\{ e_{i}\}$ is called a 
\textit{canonical basis} with respect to the initial data if
$g$ is orthonormal and $k$ is diagonal with respect to it, and if
the structure constants $\gamma^{i}_{jk}=\epsilon_{jkl}n^{li}$
associated with the basis have the property that $n$ is diagonal
and the diagonal elements $n_{i}$ fall into one of 
the classes given in Table \ref{table:bianchiA}.
\end{definition}

On pp. 3798--99 in \cite{jag3} we prove that there are canonical bases
and that the Bianchi class is well defined.

\begin{table}
\caption{Bianchi class A.}
\begin{center}
\begin{tabular}{@{}lccc}
Type & $n_{1}$ & $n_{2}$ & $n_{3}$ \\
I                   & 0 & 0 & 0 \\
II                  & + & 0 & 0 \\
V$\mathrm{I}_{0}$   & 0 & + & $-$ \\
VI$\mathrm{I}_{0}$  & 0 & + & + \\
VIII                & $-$ & + & + \\
IX                  & + & + & + \\
\end{tabular}\label{table:bianchiA}
\end{center}
\end{table}

\begin{definition}
Consider class A vacuum initial data $(G,g,k)$ of type VII${}_{0}$
or VIII and let $\{ e_{i}\}$ be a canonical basis. If $k_{2}=k_{3}$
and $n_{2}=n_{3}$, then the initial data are said to be of
\textit{Taub} type, and the corresponding basis is called a basis
of \textit{NUT} type in the Bianchi VIII case.
\end{definition}

In Lemma \ref{lemma:sim} we sort out the relation between different
canonical bases for Bianchi VIII initial data. The important point
to keep in mind here is that $e_{1}$ is well defined up to a sign.
Note also that if the initial data allow a basis of NUT type, it only
allows such bases.

Given class A initial data $(G,g,k)$ and a canonical basis 
$\{ e_{i}\}$, one can construct a globally
hyperbolic Lorenz metric $\bar{g}$ on $I\times G$ for some open
interval $I$ such that, $\mathrm{Ric}[\bar{g}]=0$, the metric induced
on $\Sigma=\{ 0\}\times G$ is $g$ and the second fundamental form
of $\Sigma$ is $k$, assuming one identifies $\Sigma$ with $G$ in
the obvious way. The metric has the form (\ref{eq:metric}), where
the $\xi^{i}$ are the duals of the canonical basis $\{ e_{i}\}$. 
By Lemma \ref{lemma:sia} the  development is independent of the
canonical basis chosen if the initial data is of Bianchi type VIII. 
Consider Bianchi VII${}_{0}$ and VIII initial
data that are not of Taub type. Then the corresponding developments are
inextendible and future causally geodesically complete, assuming one 
has chosen a suitable time orientation. These statements are proven in
\cite{jag2}. The developments constructed in \cite{jag2} will be
called \textit{class A developments}. The construction can also be
found on pp. 3798--99 of \cite{jag3}.

\begin{thm}\label{thm:bsz}
Consider Bianchi VII${}_{0}$ initial data which are not of Taub type.
Then the class A development can be written in the form
(\ref{eq:metric}), where the $\xi^{i}$ are the duals of a canonical
basis. Furthermore, there are positive constants
$\a_{i}$, $i=1,2,3$, $C$ and $T$ such that 
\begin{equation}\label{eq:att}
a_{i}(t)=\a_{i}(\ln t)^{1/2}+O(1)
\end{equation}
for $i=2,3$ and $t\geq T$. Also, there are sequences
$t_{k},s_{k}\rightarrow \infty$ and a constant $c>0$ such that
\begin{equation}\label{eq:attopt}
|a_{i}(t_{k})-\a_{i}(\ln t_{k})^{1/2}|\geq c\ \
and \ \
|a_{i}(s_{k})-\a_{i}(\ln s_{k})^{1/2}|\leq
\frac{C\ln\ln s_{k}}{(\ln s_{k})^{1/2}}
\end{equation}
for $i=2,3$. Finally
\[
a_{1}(t)=\a_{1}t[1+O(\frac{\ln\ln t}{\ln t})]
\]
for all $t\geq T$. 
\end{thm}
\textit{Remark}. The inequalities (\ref{eq:attopt}) are intended
to emphasize the optimality of (\ref{eq:att}). Observe that they
show that any lower order corrections to (\ref{eq:att}) have to 
oscillate.

The proof is to be found at the end of Section \ref{section:bsz}.
Concerning Bianchi VIII, it is of interest to contrast general 
solutions with those arising from Taub initial data. We will refer
to Bianchi VIII Taub initial data as NUT initial data.

\begin{thm}\label{thm:b8n}
Consider NUT initial data. Then the class A development can be 
written in the form (\ref{eq:metric}), where the $\xi^{i}$ are the
duals of a canonical basis. Furthermore, there are positive 
constants $\a_{1}$, $\a_{2}$ and $T$ such that
\[
a_{1}(t)=\a_{1}+O(t^{-1})\ \
and\ \
a_{2}(t)=a_{3}(t)=\a_{2}t[1+O(\frac{\ln t}{t})]
\]
for $i=2,3$ and $t\geq T$.
\end{thm}
The proof is to be found at the end of Section \ref{section:b8n}.

\begin{thm}\label{thm:b8a}
Consider Bianchi VIII initial data that are not of 
NUT type. Then the class A development can be 
written in the form (\ref{eq:metric}), where the $\xi^{i}$ are the
duals of a canonical basis. Furthermore, there are positive constants
$\a_{i}$, $i=1,2,3$, $c_{0}$ and a $T$ such that 
\[
a_{1}(t)=\a_{1}(\ln t)^{1/2}[1+
O(\frac{\ln\ln t}{\ln t})],\ \ \
a_{i}(t)=\a_{i}t[1+
O(\frac{\ln\ln t}{\ln t})]
\]
for $i=2,3$ and $t\geq T$, and
\[
\frac{a_{2}(t)}{a_{3}(t)}=c_{0}+O(t^{-1})
\]
for $t\geq T$.
\end{thm}
\textit{Remark}. By the results obtained in this article, it is
possible to obtain further terms in the expansion. However, the 
computations involved are long and have for this reason not 
been carried out.

The proof is to be found at the end of Section \ref{section:b8as}.

Let us describe the
connection between Theorem \ref{thm:b8n}, \ref{thm:b8a} and the
topology of the compactified spatial hypersurfaces, starting by
clarifying what we mean by compactifications. Consider class A
vacuum initial data $(G,g,k)$. 
We will call a diffeomorphism $\phi$ of $G$ an \textit{isometry of the 
initial data} if $\phi^{*}g=g$ and $\phi^{*}k=k$. Let $\Gamma$ be a
free and properly discontinuous group of isometries of the initial data.
We get a solution $(\tilde{g},\tilde{k})$ of Einstein's constraint 
equations on $G/\Gamma$, and if this manifold is compact, we say that 
$(G/\Gamma,\tilde{g},\tilde{k})$ is a \textit{compactification} of 
$(G,g,k)$. 

Assume now that $G$ is simply connected and let $(I\times G,\bar{g})$
be a class A development. It follows from Lemma \ref{lemma:isoa} that if
$\phi$ is an isometry of the initial data, then the diffeomorphism
$(1,\phi)$ from $I\times G$ to itself defined by
$(1,\phi)(t,h)=[t,\phi(h)]$ is an isometry of the development. 
Consequently if $\Gamma$ is a free and properly discontinuous group
of isometries of the initial data, $(1,\Gamma)$ is a free and properly
discontinuous group of isometries of the class A development. Thus
we get a vacuum Lorentz metric on $I\times G/\Gamma$ consistent with
the compactified initial data. We will refer to the corresponding
Lorentz manifold as a \textit{spatially compactified class A 
development}.

A $3$-manifold is said to be a \textit{Seifert
fibred space} if it satisfies the following two conditions:
\begin{enumerate}
\item It can be written as a disjoint union of circles.
\item Each circle fibre has an open neighbourhood $U$ satisfying:
	\begin{enumerate}
	\item $U$ can be written as a disjoint union of circle fibres.
	\item $U$ is isomorphic either to a solid torus or a cylinder
	where the ends have been identified after a rotation by a rational
	angle.
	\end{enumerate} 
\end{enumerate}
When we say that $U$ is isomorphic to a solid torus, we mean that 
$U$ is diffeomorphic to a solid torus and that the circle fibres of
$U$ are mapped to the natural circle fibres of the solid torus. 
Note that there are different definitions of Seifert fibred
spaces in the literature. In particular, our definition coincides with
the original definition by Seifert but not with that of Scott
\cite{scott}.

\begin{thm}\label{thm:seifert}
Consider Bianchi VIII initial data $(G,g,k)$ where $G$ is simply
connected and assume that $\Gamma$ is a free and properly
discontinuous group of isometries of the initial data such that 
$G/\Gamma$ is compact. Then $G/\Gamma$ is a Seifert fibred space,
and if $e_{1}$ is the first element in a canonical basis for the 
initial data, the integral curves of $e_{1}$ map to the circle fibres.
As noted above, the action of $\Gamma$ extends to the class A 
development, so that given a circle fibre we can speak of its length
$l(t)$ at time $t$. By Theorem \ref{thm:b8n} we conclude that if
the initial data are of NUT type, there is an $l_{0}>0$ such that 
\[
l(t)=l_{0}+O(t^{-1})
\]
and by Theorem \ref{thm:b8a} we conclude that if the initial data are
not of NUT type, there is a $c_{0}>0$ such that 
\[
l(t)=c_{0}(\ln t)^{1/2}[1+O(\frac{\ln\ln t}{\ln t})].
\]
\end{thm}
\textit{Remark}. For any initial data there are $\Gamma$ such that
$G/\Gamma$ is compact. In fact, let $\tsl$ denote the universal 
covering group of $\sl$, which is a simply connected Lie group of
Bianchi type VIII. Then, for any higher genus surface
$\Sigma_{p}$ with hyperbolic metric $h$, there is a subgroup $\Xi_{p}$
of $\tsl$ such that $\Xi_{p}$ acting on the left is a free and
properly discontinuous group of diffeomorphims with $\tsl/\Xi_{p}$
diffeomorphic to the unit tangent bundle of $\Sigma_{p}$ with
respect to the metric $h$. This is of course well known, but
we include a proof in Lemma \ref{lemma:ce} since there seems
to be some confusion in the GR literature.

The proof is to be found at the very end of the article.

The associated question of what the possible topologies are for
compactifications of initial data is not addressed here. If the
initial data are of NUT type, the question has been answered,
see \cite{scott} and references therein, since the isometry group for
NUT initial data is the same as the isometry group for the Thurston
geometry on $\tsl$. Furthermore, if $\Gamma$ consists only of left
translations, what the possible topologies are has been sorted out
in \cite{ray}. However, the question of greatest interest when
discussing Bianchi VIII is what the possible topologies are when
$\Gamma$ is a subgroup of the isometry group of general initial data.
This case falls between the cases that have been handled, cf. 
Lemma \ref{lemma:isoa}.

Let us try to describe what makes the analysis possible in the 
expanding direction of Bianchi VIII vacuum spacetimes. 
For a definition of the variables of Wainwright and Hsu, see Section
\ref{section:whsu}. It turns out that for general, i.e. non NUT, 
Bianchi VIII solutions,
\[
h=\Sm^{2}+\frac{3}{4}(\Nt-\Nth)^{2}=\frac{1}{4\tau}+
O(\frac{\ln\tau}{\tau^{2}}),
\]
see (\ref{eq:has2}), where the first equality is a definition,  but that 
\[
(\Sm')^{2}+\frac{3}{4}(\Nt'-\Nth')^{2}=
c\tau^{-5/2}e^{3\tau}[1+O(\frac{\ln\tau}{\tau})],
\]
where $c>0$ is a constant, which follows from (\ref{eq:txpr}),
(\ref{eq:typr}), (\ref{eq:ntas}) and (\ref{eq:has2}). 
This is a quantitative illustration of the oscillatory behaviour of
$\Sm$ and $\Nt-\Nth$, which we give more detailed description of
in Section \ref{section:appr}. On the other hand, $\Spp$ and 
$\No(\Nt+\Nth)$ have bounded derivatives. Interestingly enough,
the fact that some variables have bounded derivatives
and some have derivatives that tend to infinity exponentially
is what makes the problem manageable analytically. By averaging over
the variables with bounded derivatives during a period of the
oscillation in the rapidly varying variables, one obtains an
iteration. Since the frequency of the oscillations tends to infinity
exponentially with time, this approximation is very good. In fact,
one can obtain series expansions of the form
\[
\Spp(\tau)=\frac{1}{2}-\frac{1}{4\tau}+\frac{\ln\tau}{8\tau^2}+
\frac{c_{u}}{\tau^{2}}-\frac{\ln^{2}\tau}{16\tau^{3}}+
O(\frac{\ln\tau}{\tau^{3}}),
\]
see (\ref{eq:uestf}), where $u=\Spp-1/2$ and $c_{u}$ is some constant. 
One can also obtain similar expansions for the other slowly varying
variables, and there seems to be no reason to believe that one could 
not obtain further terms in the expansions. Note that heuristic
arguments supporting expansions of this form are provided in
\cite{wain}. This article furthermore deals with the case when matter
is present.

The article is structured as follows. In Section \ref{section:whsu}
we describe the variables of Wainwright and Hsu and in Section
\ref{section:back} we give the necessary background from \cite{jag3}.
In Sections \ref{section:bsz} and \ref{section:b8n} we then analyze Bianchi
$\mathrm{VII}_{0}$ and Bianchi VIII NUT. The heart of the paper
then consists of Sections \ref{section:appr} and \ref{section:b8as},
which contain the analysis of general Bianchi VIII solutions. 
The remaining sections contain the material necessary to connect
the asymptotic expansions of the metric with the Seifert fibred
structure of the compactified spatial hypersurfaces. The main point
is of course to identify the $e_{1}$ appearing in a canonical basis
with the fibre direction in the Seifert fibration. We do this
in a very explicit way, deriving the isometry group for the relevant
initial data, and proving that the cocompact subgroups of the isometry
group have to have a particular structure. While doing this, we always
keep track of $e_{1}$ in order to achieve the desired objective. The 
argument to prove that the compactifications are Seifert fibred is 
taken from \cite{thurston}. In fact most of the arguments presented
are available in the literature. However, they are spread out and are
not always so accessible to outsiders. For this reason the last two 
chapters also have a pedagogical purpose. We wish to carry out the 
necessary proofs without assuming more of the reader than basic
differential geometry, Lie group theory and knowledge of covering
spaces.

\section{The equations of Wainwright and Hsu}\label{section:whsu}
Here we formulate the equations we will use and state some 
properties. The equations were obtained in \cite{whsu} and are based
on a formulation given in \cite{emac}. A brief derivation can
be found in \cite{jag3}. 
For Bianchi class A spacetimes, Einstein's vacuum
equations take the following form in the formulation due to Wainwright
and Hsu
\begin{eqnarray}
\No'&=&(q-4\Spp)\No  \nonumber \\
\Nt'&= &(q+2\Spp +2\sqrt{3}\Sm)\Nt  \nonumber \\
\Nth'&=& (q+2\Spp -2\sqrt{3}\Sm)\Nth  \label{eq:whsu}\\
\Spp'&=& -(2-q)\Spp-3S_{+}  \nonumber \\
\Sm'&=& -(2-q)\Sm-3S_{-} \nonumber 
\end{eqnarray}
where the prime denotes derivative with respect to $\tau$ and
\begin{eqnarray}
q & = & 2(\Spp^2+\Sm^2) \nonumber \\
S_{+} & = & \frac{1}{2}[(\Nt-\Nth)^2-\No(2\No-\Nt-\Nth)]
\label{eq:whsudef}\\
S_{-} & = & \frac{\sqrt{3}}{2}(\Nth-\Nt)(\No-\Nt-\Nth). \nonumber
\end{eqnarray}
The vacuum Hamiltonian constraint is
\begin{equation}
\Spp^2+\Sm^2+\frac{3}{4}[\No^2+\Nt^2+\Nth^2-2(\No\Nt+\Nt\Nth+\No\Nth)]=1.
\label{eq:constraint}
\end{equation}
The above equations have certain
symmetries described in \cite{whsu}. By permuting
$\No,\Nt,\Nth$ arbitrarily  we get new solutions if we at the same 
time carry out appropriate
combinations of rotations by integer multiples of $2\pi/3$
and reflections in the $(\Spp,\Sm)$-plane.
Changing the sign of all the $N_{i}$ at the same time does not change
the equations. Since the sets 
$N_{i}>0$, $N_{i}<0$ and $N_{i}=0$ are invariant under the flow of 
the equation we may classify solutions to 
(\ref{eq:whsu})-(\ref{eq:constraint}) accordingly. Taking the
symmetries into account, we get Table \ref{table:bianchiA}.
When we speak of Bianchi VIII solutions we will assume that $\Nt,\
\Nth>0$ and that $\No<0$. 
We only consider solutions to (\ref{eq:whsu})-(\ref{eq:constraint})
which are not of Bianchi IX type. 
By the constraint (\ref{eq:constraint}) we conclude that $q\leq
2$ for such solutions. As a consequence the $N_{i}$ cannot 
grow faster than exponentially due to (\ref{eq:whsu}). The 
solutions we consider can thus not blow up in a finite time
so that we have existence intervals of the form $(-\infty,\infty)$.
The set $\Sm=0$, $\Nt=\Nth$ is invariant under the flow of
(\ref{eq:whsu})-(\ref{eq:constraint}). Applying the symmetries to 
this set we get new invariant sets. Solutions which are contained
in this invariant set are said to be of \textit{Taub} type.

\section{Background}\label{section:back}

This article is a continuation of the analysis presented in
\cite{jag3}, and for the benefit of the reader, we wish to 
restate the terminology and main conclusions of that paper. Let
us first note that the trace of the second fundamental form of the 
spatial hypersurfaces of homogeneity $\theta$ satisfies
\begin{equation}\label{eq:tp}
\theta'=-(1+q)\theta.
\end{equation}
This is equation (17) of \cite{jag3}. Observe that it is decoupled from
(\ref{eq:whsu})-(\ref{eq:constraint}). The prime is with respect to 
the same $\tau$-time as in (\ref{eq:whsu})-(\ref{eq:constraint}).
This time is related to the $t$-time in (\ref{eq:metric}) by 
\begin{equation}\label{eq:ttau}
\frac{dt}{d\tau}=\frac{3}{\theta},
\end{equation}
which is equation (16) of \cite{jag3}. Finally, the $a_{i}$
appearing in (\ref{eq:metric}) satisfy
\begin{equation}\label{eq:aitau}
a_{i}(\tau)=\exp[\int_{0}^{\tau}(3\Sigma_{i}+1)d s],
\end{equation}
where
\begin{equation}\label{eq:sigi}
\Sigma_{1}=-\frac{2}{3}\Spp,\ 
\Sigma_{2}=\frac{1}{3}\Spp+\frac{1}{\sqrt{3}}\Sm,\
\Sigma_{3}=\frac{1}{3}\Spp-\frac{1}{\sqrt{3}}\Sm.
\end{equation}
These are equations (25) and (26) of \cite{jag3}. We will quite
consistently abuse notation and write $a_{i}(\tau)$ when we mean
$a_{i}[t(\tau)]$ etc. However, when we talk of $N_{i}$, $\Spp$ and
$\Sm$, we will always stick to the time associated with the equations
(\ref{eq:whsu})-(\ref{eq:constraint}).
Let us now state what we need to know 
concerning Bianchi $\mathrm{VII}_{0}$. 
The main theorem concerning Bianchi $\mathrm{VII}_{0}$ solutions
is the following.
\begin{thm}\label{thm:msz}
Consider a non-Taub Bianchi VII${}_{0}$ solution to 
(\ref{eq:whsu})-(\ref{eq:constraint}) with $\No=0$ and 
$\Nt,\ \Nth>0$. Then there is an $n_{0}>0$ such that 
\[
\lim_{\tau\rightarrow\infty}(\Spp,\Sm,\Nt,\Nth)=(-1,0,n_{0},n_{0}).
\]
\end{thm}
This is Theorem 1.1 of \cite{jag3}. Note the problem that arises
when one wants to use this information to estimate the behaviour
of $a_{2}$ and $a_{3}$; the relevant integrand tends to
zero. When considering Bianchi VII${}_{0}$ solutions, we will always
assume that  $\Nt,\ \Nth>0$
and $\No=0$. We will also only be interested in non-Taub type solutions.
Thus, $\Sm^{2}+(\Nt-\Nth)^{2}>0$ for all times. This implies
that $-1<\Spp<1$ for all times. Thus we can define the smooth
functions
\begin{eqnarray}
x & = & \frac{\Sm}{(1-\Spp^{2})^{1/2}}\label{eq:xdef1}\\
y & = & \frac{\sqrt{3}}{2}\frac{\Nt-\Nth}{(1-\Spp^{2})^{1/2}}.
\label{eq:ydef1}
\end{eqnarray}
Let
\begin{equation}\label{eq:gdef}
\tg=3(\Nt+\Nth)+2(1+\Spp)xy.
\end{equation}
Then $x'=-\tg y$ and $y'=\tg x$. By the constraint, $x^2+y^2=1$, so that
we can choose a $\phi_{0}$ such that $[x(\tau_{0}),y(\tau_{0})]=
[\cos (\phi_{0}),\sin (\phi_{0})]$. Define
\begin{equation}\label{eq:xidef1}
\xi(\tau)=\int_{\tau_{0}}^{\tau}\tg (s) d s+\phi_{0}.
\end{equation}
Then $x(\tau)=\cos[\xi(\tau)]$ and $y(\tau)=\sin [\xi(\tau)]$.
Observe that if one combines Theorem \ref{thm:msz}
with (\ref{eq:gdef}), one gets the conclusion that $\tg\rightarrow
6n_{0}$. This means that $x$ and $y$ will oscillate forever and that
they in the limit will behave more and more like $\cos(6n_{0}\tau)$
and $\sin(6n_{0}\tau)$. For future reference, let us just note that 
there is a $\tau_{0}$ such that
\begin{equation}\label{eq:glb}
\tg(\tau)\geq 5n_{0}
\end{equation}
for all $\tau\geq \tau_{0}$. We will need Lemma 7.2 and 7.3 of \cite{jag3},
which we now state.
\begin{lemma}
Consider a non-Taub Bianchi VII${}_{0}$ solution to 
(\ref{eq:whsu})-(\ref{eq:constraint}) with $\Nt,\Nth> 0$,
and let $\tau_{0}$ be such that (\ref{eq:glb}) is fulfilled for
$\tau\geq \tau_{0}$. Let $\tau_{0}\leq \tau_{a}< \tau_{b}$, and assume
\[
\xi(\tau_{b})-\xi(\tau_{a})=\pi.
\]
Then there is a $T$ such that 
\begin{equation}\label{eq:spvar3}
|1-\frac{(1+\Spp)(\tau_{1})}
{(1+\Spp)(\tau_{2})}|\leq C
\frac{(1+\Spp)(\tau_{b})}{(\Nt+\Nth)(\tau_{\max})}
\end{equation}
and
\begin{equation}\label{eq:nvar3}
|1-\frac{(\Nt+\Nth)(\tau_{1})}
{(\Nt+\Nth)(\tau_{2})}|\leq C
\frac{(1+\Spp)(\tau_{b})}{(\Nt+\Nth)(\tau_{\max})}
\end{equation}
if $\tau_{a}\geq T$, where $\tau_{\max}$ yields the maximum value
of $\Nt+\Nth$ in $[\tau_{a},\tau_{b}]$ and 
$\tau_{1},\tau_{2}$ are arbitrary elements of $[\tau_{a},\tau_{b}]$. 
The constant $C$ only depends on 
the constant $5n_{0}$ appearing in (\ref{eq:glb}).
\end{lemma}
\begin{lemma}\label{lemma:oscvar}
Consider a non-Taub Bianchi VII${}_{0}$ solution to 
(\ref{eq:whsu})-(\ref{eq:constraint}) with $\Nt,\Nth> 0$,
and let $\tau_{0}$ be such that (\ref{eq:glb}) is fulfilled
for $\tau\geq \tau_{0}$. 
Let $\tau_{0}\leq \tau_{a}< \tau_{b}$, and assume
\[
\xi(\tau_{b})-\xi(\tau_{a})=\pi.
\]
Then there is a $T$ such that 
\begin{equation}\label{eq:oscvar}
|\int_{\tau_{a}}^{\tau_{b}}[\Sm^2-(1+\Spp)]d s|\leq
C\int_{\tau_{a}}^{\tau_{b}}(1+\Spp)^2 d s
\end{equation}
if $\tau_{a}\geq T$, for some constant $C$ depending only on the
constant $5n_{0}$ appearing in (\ref{eq:glb}).
\end{lemma}
Note that it is far from obvious that a statement such as
(\ref{eq:oscvar}) should hold.
The integrand is of the order of magnitude $1+\Spp$, but the average
over a period is of the order of magnitude $(1+\Spp)^{2}$. Finally,
note that by the proof of Lemma 7.5 in \cite{jag3}, there
is for every $0<\a<1$ a $0<c_{\a}<\infty$ and a $T_{\a}$ such that
\begin{equation}\label{eq:spprel}
|1+\Spp(\tau)|\leq\frac{c_{\a}}{\tau^{\a}}
\end{equation}
for all $\tau\geq T_{\a}$.

Concerning Bianchi VIII, we will only need to know the contents of 
Theorem 1.2 of \cite{jag3}. 
\begin{thm}\label{thm:b8}
Let $(\Spp,\Sm,\No,\Nt,\Nth)$ be a Bianchi VIII solution to
(\ref{eq:whsu})-(\ref{eq:constraint}) with $\No<0$ and 
$\Nt,\Nth>0$. Then
\[
\lim_{\tau\rightarrow \infty}\No=0,\
\lim_{\tau\rightarrow \infty}\Nt=\infty,\
\lim_{\tau\rightarrow \infty}\Nth=\infty.
\]
Furthermore
\[
\lim_{\tau\rightarrow \infty}\Spp=\frac{1}{2},\
\lim_{\tau\rightarrow \infty}\Sm=0
\]
and 
\[
\lim_{\tau\rightarrow \infty}\No(\Nt+\Nth)=-\frac{1}{2},\
\lim_{\tau\rightarrow \infty}(\Nt-\Nth)=0.
\]
\end{thm}

\section{Bianchi VII${}_{0}$}\label{section:bsz}

We start where \cite{jag3} ended.

\begin{prop}\label{prop:spas}
Consider a non-Taub Bianchi VII${}_{0}$ solution to 
(\ref{eq:whsu})-(\ref{eq:constraint}). Then there is a 
$T$ and a $C$ such that 
\begin{equation}\label{eq:spest1}
|1+\Spp(\tau)-\frac{1}{2\tau}|\leq\frac{C\ln\tau}{\tau^2}
\end{equation}
for all $\tau\geq T$.
\end{prop}

\textit{Proof}. We have
\[
(1+\Spp)'=-(2-q)(1+\Spp)
\]
whence
\begin{equation}\label{eq:oot}
\frac{(1+\Spp)'}{(1+\Spp)^2}=-2-\frac{(2-q)-2(1+\Spp)}{1+\Spp}.
\end{equation}
Let $\xi$ be defined by (\ref{eq:xidef1}) and assume
$[\tau_{a},\tau_{b}]$ is an interval such that
\[
\xi(\tau_{b})-\xi(\tau_{a})=\pi.
\]
By (\ref{eq:spvar3}) and the fact that $\Nt+\Nth$ is bounded
from below by a positive constant, we deduce the existence of a
constant $C$ such that if $\tau_{a}$ is big enough, then
\[
|\frac{1}{1+\Spp(\tau_{1})}-\frac{1}{1+\Spp(\tau_{2})}|\leq C
\]
for all $\tau_{1},\tau_{2}\in [\tau_{a},\tau_{b}]$. Thus
\[
\int_{\tau_{a}}^{\tau_{b}}\frac{(2-q)-2(1+\Spp)}{1+\Spp}ds=
\frac{1}{1+\Spp(\tau_{a})}
\int_{\tau_{a}}^{\tau_{b}}[(2-q)-2(1+\Spp)]ds+\epsilon_{1}
\]
where
\[
|\epsilon_{1}|\leq C\int_{\tau_{a}}^{\tau_{b}}[1+\Spp(\tau)]d\tau.
\]
Observe now that by Lemma \ref{lemma:oscvar},
\[
|\int_{\tau_{a}}^{\tau_{b}}[\Sm^2-(1+\Spp)]d\tau|\leq
C\int_{\tau_{a}}^{\tau_{b}}[1+\Spp(\tau)]^{2}d\tau.
\]
Since 
\begin{equation}\label{eq:ref}
(2-q)-2(1+\Spp)=2(1+\Spp)-2\Sm^2-2(1+\Spp)^2,
\end{equation}
we conclude that 
\[
|\int_{\tau_{a}}^{\tau_{b}}[(2-q)-2(1+\Spp)]d\tau|\leq
C\int_{\tau_{a}}^{\tau_{b}}[1+\Spp(\tau)]^{2}d\tau\leq
\]
\[
\leq C[1+\Spp(\tau_{a})]\int_{\tau_{a}}^{\tau_{b}}[1+\Spp(\tau)]d\tau.
\]
Thus
\begin{equation}\label{eq:diffest}
|\int_{\tau_{a}}^{\tau_{b}}\frac{(2-q)-2(1+\Spp)}{1+\Spp}ds|
\leq C\int_{\tau_{a}}^{\tau_{b}}[1+\Spp(\tau)]d\tau.
\end{equation}
Since 
\[
|\int_{\tau_{1}}^{\tau_{2}}\frac{(2-q)-2(1+\Spp)}{1+\Spp}ds|\leq C
\]
if $[\tau_{1},\tau_{2}]$ corresponds to less than one multiple of
$\pi$, we conclude that 
\begin{equation}\label{eq:wth}
|\frac{1}{1+\Spp(0)}-\frac{1}{1+\Spp(\tau)}+2\tau|\leq
C+C\int_{0}^{\tau}[1+\Spp(s)]ds
\end{equation}
for $\tau\geq 0$ where we have also used (\ref{eq:oot}) and
(\ref{eq:diffest}). Observe that the estimate (\ref{eq:diffest})
may only be applicable for $\tau\geq T$, but as this is a bounded
time, the constant takes care of the discrepancy. By 
(\ref{eq:spprel}) we conclude that for every $\a>0$, there is 
a constant $C_{\a}$ such that 
\[
|-\frac{1}{1+\Spp(\tau)}+2\tau|\leq
C_{\a}\tau^{\a}
\]
for all $\tau\geq 1$. Inserting this information into (\ref{eq:wth})
we get the conclusion that 
\[
|-\frac{1}{1+\Spp(\tau)}+2\tau|\leq
C\ln\tau
\]
for all $\tau\geq 2$. This proves the proposition. $\hfill\Box$

\begin{cor}\label{cor:spint}
Consider a non-Taub Bianchi VII${}_{0}$ solution to 
(\ref{eq:whsu})-(\ref{eq:constraint}). There are constants $c_{s+}$
and $C$ such that 
\[
|\int_{1}^{\tau}[1+\Spp(s)]ds-\frac{1}{2}\ln\tau-c_{s+}|\leq
\frac{C\ln\tau}{\tau}
\]
for all $\tau\geq 2$.
\end{cor}

\textit{Proof}. Let
\[
c_{s+}=\int_{1}^{\infty}[1+\Spp(\tau)-\frac{1}{2\tau}]d\tau,
\]
which is convergent due to the previous proposition. The estimate
(\ref{eq:spest1}) yields the conclusion of the corollary. $\hfill\Box$

\begin{cor}\label{cor:q}
Consider a non-Taub Bianchi VII${}_{0}$ solution to 
(\ref{eq:whsu})-(\ref{eq:constraint}). There are constants $c_{q}$
and $C$ such that 
\[
|\int_{1}^{\tau}(2-q)d\tau-\ln\tau-c_{q}|\leq
\frac{C\ln\tau}{\tau}
\]
for all $\tau\geq 2$.
\end{cor}

\textit{Proof}. Observe that by the proof of Proposition
\ref{prop:spas},
\[
\int_{1}^{\tau}[(2-q)-2(1+\Spp)]ds
\]
converges as $\tau\rightarrow \infty$. Let $\a$ be the limit. Then
Lemma \ref{lemma:oscvar} and (\ref{eq:ref}) yield
\[
|\int_{1}^{\tau}[(2-q)-2(1+\Spp)]ds-\a|\leq C\int_{\tau}^{\infty}
[1+\Spp(s)]^{2}\leq
\frac{C}{\tau},
\]
if $\tau$ is big enough, where we have also used (\ref{eq:spest1}).
The conclusion now follows from Corollary \ref{cor:spint}. $\hfill\Box$

\begin{cor}
Consider a non-Taub Bianchi VII${}_{0}$ solution to 
(\ref{eq:whsu})-(\ref{eq:constraint}). There is a constant $C$
such that
\begin{equation}\label{eq:nqest}
|\frac{\Nt}{\Nth}-1|\leq\frac{C}{\tau^{1/2}}
\end{equation}
for all $\tau\geq 1$. Furthermore, there are sequences $\tau_{k},
\hat{\tau}_{k}\rightarrow\infty$ and a constant $c>0$ such that
\begin{equation}\label{eq:nqlb}
|\frac{\Nt(\tau_{k})}{\Nth(\tau_{k})}-1|\geq\frac{c}{\tau_{k}^{1/2}}\
 \ \mathrm{and}\ \
|\frac{\Nt(\hat{\tau}_{k})}{\Nth(\hat{\tau}_{k})}-1|=0.
\end{equation}
\end{cor}
\textit{Remark}. The equation (\ref{eq:nqlb}) is merely intended to 
emphasize the optimality of (\ref{eq:nqest}).

\textit{Proof}. By the constraint,
\[
\frac{3}{2}(\Nt-\Nth)^{2}=2-q\leq 2(1-\Spp^2)=2(1-\Spp)(1+\Spp),
\]
which, when combined with (\ref{eq:spest1}) and the fact that 
$\Nth$ is bounded from below by a positive constant yields 
(\ref{eq:nqest}). Due to the oscillatory behaviour of the solutions
noted in the paragraph preceeding (\ref{eq:glb}), we
conclude the existence of time sequences $\tau_{k},
\hat{\tau}_{k}\rightarrow\infty$ such that $\Sm(\tau_{k})=0$ and
$(\Nt-\Nth)(\hat{\tau}_{k})=0$. Since $\Nth$ is bounded from below by a 
positive constant, we conclude that the second part of (\ref{eq:nqlb}) 
holds. Finally, the first part of (\ref{eq:nqlb})
follows from (\ref{eq:spest1}), the constraint, $\Sm(\tau_{k})=0$ and
the fact that $\Nth$ is bounded from below by a positive constant.
$\hfill\Box$

\begin{cor}\label{cor:psiest}
Consider a non-Taub Bianchi VII${}_{0}$ solution to 
(\ref{eq:whsu})-(\ref{eq:constraint}). Then there is a constant
$C$ and a $T$ such that
\[
|\psi(\tau)|\leq\frac{C}{\tau^{1/2}}
\]
for all $\tau\geq T$, where
\[
\psi(\tau)=\ln\frac{\Nt(0)}{\Nth(0)}+\int_{0}^{\tau}4\sqrt{3}\Sm ds.
\] 
Furthermore, there are sequences $\tau_{k},
\hat{\tau}_{k}\rightarrow\infty$ and a constant $c>0$ such that
\[
|\psi(\tau_{k})|\geq\frac{c}{\tau_{k}^{1/2}}\ \
\mathrm{and}\ \
\psi(\hat{\tau}_{k})=0.
\]
\end{cor}

\textit{Proof}. Observe that 
\[
\frac{\Nt(\tau)}{\Nth(\tau)}=e^{\psi(\tau)}.
\]
The statements thus follow from the previous corollary. $\hfill\Box$

\begin{prop}\label{prop:aiw}
Consider a non-Taub Bianchi VII${}_{0}$ solution to 
(\ref{eq:whsu})-(\ref{eq:constraint}). There are positive constants 
$\a_{i}$, $i=1,2,3$, $C$ and a $T$ such that 
\begin{equation}\label{eq:atest}
|a_{i}(\tau)-\a_{i}\tau^{1/2}|\leq C
\end{equation}
for $i=2,3$ and all $\tau\geq T$. Furthermore, there are time sequences
$\tau_{k},\ \hat{\tau}_{k}\rightarrow\infty$ and a $c>0$ such that 
\[
|a_{i}(\tau_{k})-\a_{i}\tau_{k}^{1/2}|\geq c\ \
and\ \
|a_{i}(\hat{\tau}_{k})-\a_{i}\hat{\tau}_{k}^{1/2}|\leq 
\frac{C\ln \hat{\tau}_{k}}{\hat{\tau}_{k}^{1/2}}
\]
for $i=2,3$. Finally
\begin{equation}\label{eq:aoestw}
|\tau a_{1}(\tau)e^{-3\tau}-\a_{1}|\leq \frac{C\ln\tau}{\tau}
\end{equation}
for all $\tau\geq T$. 
\end{prop}

\textit{Proof}. By Corollary \ref{cor:spint} and \ref{cor:psiest},
there is a constant $c_{s,2}$ such that 
\[
\kappa(\tau)=\int_{0}^{\tau}(1+\Spp+\sqrt{3}\Sm)ds
-\frac{1}{2}\ln\tau-c_{s,2}=O(\tau^{-1/2})
\]
for $\tau$ big enough, where the first equality defines 
$\kappa$. Furthermore, we
conclude the existence of a time sequence $\tau_{k}\rightarrow\infty$
and a $c>0$ such that 
\[
|\kappa(\tau_{k})|\geq 
\frac{c}{\tau_{k}^{1/2}}.
\]
There is also a time sequence $\hat{\tau}_{k}\rightarrow\infty$ such that 
\[
|\kappa(\hat{\tau}_{k})|\leq\frac{C\ln\hat{\tau}_{k}}{\hat{\tau}_{k}}.
\]
Since
\[
|a_{2}(\tau)-e^{c_{s,2}}\tau^{1/2}|=
e^{c_{s,2}}\tau^{1/2}|e^{\kappa(\tau)}-1|,
\]
the statements concerning $a_{2}$ follow. The proof for
$a_{3}$ is similar, and one can see that the time sequences
$\tau_{k}$ and $\hat{\tau}_{k}$ can be taken to be the same for $a_{2}$ and 
$a_{3}$. Since 
\[
a_{1}(\tau)=\exp[-\int_{0}^{\tau}(2\Spp-1)]=
\exp[3\tau-2\int_{0}^{\tau}(1+\Spp)ds],
\]
the statement concerning $a_{1}$ follows from Corollary
\ref{cor:spint}. $\hfill\Box$

\begin{lemma}
Consider a non-Taub Bianchi VII${}_{0}$ solution to 
(\ref{eq:whsu})-(\ref{eq:constraint}). Then there are constants
$C$ and $c_{t}>0$ and a $T$ such that
\begin{equation}\label{eq:ttrel}
|\tau\cdot t(\tau)e^{-3\tau}-c_{t}|\leq\frac{C\ln\tau}{\tau}
\end{equation}
for all $\tau\geq T$.
\end{lemma}

\textit{Proof}. Since (\ref{eq:tp}) and Corollary \ref{cor:q} hold, we
conclude that there is a constant $c_{\theta}>0$ such that 
\[
|\frac{\tau e^{-3\tau}}{\theta(\tau)}-c_{\theta}|\leq
\frac{C\ln\tau}{\tau}
\]
for $\tau$ big enough. Since we have (\ref{eq:ttau}),
it is of interest to observe that 
\[
|\tau
 e^{-3\tau}\int_{1}^{\tau}\frac{3}{s}e^{3s}ds-1|\leq\frac{C}{\tau}
\]
and that 
\[
|\tau e^{-3\tau}\int_{1}^{\tau}\frac{\ln s}{s^{2}}e^{3s}ds|
\leq\frac{C\ln\tau}{\tau}.
\]
These observations together yield the statement of the lemma. $\hfill\Box$

\textit{Proof of Theorem \ref{thm:bsz}}.
Equation (\ref{eq:ttrel}) implies that 
\[
\ln t=\ln c_{t}-\ln \tau+3\tau+\ln[1+\zeta(\tau)]
\]
where 
\[
\zeta(\tau)=\frac{1}{c_{t}}[\tau\cdot t(\tau)e^{-3\tau}-c_{t}].
\]
We thus get
\[
3\tau=\ln t+\phi(\tau),
\]
where $|\phi(\tau)|\leq C\ln\tau$. In consequence,
\[
\tau=\frac{1}{3}\ln t+\rho(t),
\]
where $|\rho(t)|\leq C\ln\ln t$. Thus
\[
\tau^{1/2}=(\frac{1}{3}\ln t+\rho(t))^{1/2}=
\frac{1}{\sqrt{3}}(\ln t)^{1/2}+
\frac{1}{\sqrt{3}}(\ln t)^{1/2}[(1+\frac{3 \rho(t)}{\ln t})^{1/2}-1].
\]
Consequently,
\[
|\tau^{1/2}-\frac{1}{\sqrt{3}}(\ln t)^{1/2}|\leq 
\frac{C\ln\ln t}{(\ln t)^{1/2}}.
\]
Combining these observations with Proposition \ref{prop:aiw}, we get the
conclusions of the theorem as far as $a_{2}$ and $a_{3}$ are
concerned, assuming we rename $\a_{2}$ and $\a_{3}$. Estimate
\[
|\frac{a_{1}(t)}{t}-\frac{\a_{1}}{c_{t}}|=
\frac{1}{\tau\cdot t(\tau)}e^{3\tau}|\tau a_{1}(\tau)e^{-3\tau}
-\a_{1}+\a_{1}-\frac{\a_{1}}{c_{t}}\tau\cdot t(\tau)e^{-3\tau}|
\leq \frac{C\ln\ln t}{\ln t}
\]
where we have used (\ref{eq:aoestw}) and (\ref{eq:ttrel}).
If we rename $\alpha_{1}/c_{t}$ to $\alpha_{1}$, the theorem 
follows. $\hfill\Box$

\section{Bianchi VIII NUT}\label{section:b8n}

Let us consider Bianchi VIII solutions to
(\ref{eq:whsu})-(\ref{eq:constraint}) such that $\Nt=\Nth$ and
$\Sm=0$. 

\begin{prop}\label{prop:b8na}
Consider a Bianchi VIII NUT solution to 
(\ref{eq:whsu})-(\ref{eq:constraint}). Then there are constants
$C$ and $c_{N}>0$ and a $T$ such that 
\begin{equation}\label{eq:spest}
|\Spp(\tau)-\frac{1}{2}|\leq Ce^{-3\tau/2},
\end{equation}
\begin{equation}\label{eq:netest}
|(\No\Nt)(\tau)+\frac{1}{4}|\leq Ce^{-3\tau/2}
\end{equation}
and
\begin{equation}\label{eq:nest}
|\Nt-c_{N}e^{3\tau/2}|\leq C
\end{equation}
for all $\tau\geq T$.
\end{prop}

\textit{Proof}. Let us reformulate the expression for $\Spp'$.
Using the constraint (\ref{eq:constraint}), we can express $\No\Nt$
in terms of the other variables in order to obtain
\[
\Spp'=-2(1-\Spp^2)\Spp-2(1-\Spp^2)(-\frac{1}{2})+\frac{9}{4}\No^2=
-2(1-\Spp^2)(\Spp-\frac{1}{2})+\frac{9}{4}\No^2.
\]
Letting $\hat{g}=\Spp-\frac{1}{2}$, $f=2(1-\Spp^2)$ and $h=9\No^2/4$, we get
the conclusion that 
\begin{equation}\label{eq:ghp}
\hat{g}'=-f\hat{g}+h.
\end{equation}
By Theorem \ref{thm:b8} we know that
$\Spp\rightarrow\frac{1}{2}$. By (\ref{eq:whsu}) we have
\[
|h(\tau)|=\frac{9}{4}\No^{2}(\tau)\leq Ce^{-5\tau/2}
\]
for $\tau\geq 0$. Since $f\rightarrow\frac{3}{2}$, we conclude that
\[
|h(\tau)\exp[\int_{0}^{\tau}f(s)ds]|\leq Ce^{-\tau/2}
\]
for all $\tau\geq 0$. By (\ref{eq:ghp}),
\[
\left(\exp[\int_{0}^{\tau}f(s)ds]\hat{g}\right)'=
h\exp[\int_{0}^{\tau}f(s)ds],
\]
so that 
\begin{equation}\label{eq:hgest}
|\hat{g}(\tau)|\leq C\exp[-\int_{0}^{\tau}f(s)ds].
\end{equation}
As a preliminary result, this yields 
\[
|\Spp(\tau)-\frac{1}{2}|\leq Ce^{-\tau},
\]
so that
\[
|\int_{0}^{\tau}f(s)ds-3\tau/2|=
|\int_{0}^{\tau}2(\frac{1}{4}-\Spp^2)ds|\leq
C\int_{0}^{\tau}e^{-s}ds\leq C,
\]
which, together with (\ref{eq:hgest})
proves (\ref{eq:spest}). By (\ref{eq:whsu}), we have
\[
(\No\Nt)(\tau)=(\No\Nt)(0)\exp[\int_{0}^{\tau}4\Spp(\Spp-\frac{1}{2})ds]
=
\]
\[
=-\frac{1}{4}\exp[-\int_{\tau}^{\infty}4\Spp(\Spp-\frac{1}{2})ds],
\]
where we have used the fact that $\No\Nt\rightarrow -1/4$, cf. Theorem
\ref{thm:b8}. Thus
\[
|(\No\Nt)(\tau)+\frac{1}{4}|\leq C\int_{\tau}^{\infty}|\Spp-\frac{1}{2}|ds
\]
proving (\ref{eq:netest}). Since 
\[
e^{-3\tau/2}\Nt(\tau)
=\Nt(0)\exp[\int_{0}^{\tau}2(\Spp+\frac{3}{2})(\Spp-\frac{1}{2})ds]
\]
we conclude the existence of a constant $c_{N}$ such that
\[
|e^{-3\tau/2}\Nt(\tau)-c_{N}|\rightarrow 0.
\]
In consequence, 
\[
|e^{-3\tau/2}\Nt(\tau)-c_{N}|=
c_{N}|\exp[-\int_{\tau}^{\infty}2(\Spp+\frac{3}{2})(\Spp-\frac{1}{2})ds]-1|
\]
and (\ref{eq:nest}) follows. $\hfill\Box$

\begin{cor}
Consider a Bianchi VIII NUT solution to
(\ref{eq:whsu})-(\ref{eq:constraint}). Then there are positive 
constants $\a_{1}$, $\a_{2}$ and $C$ such that
\[
|a_{1}(\tau)-\a_{1}|\leq Ce^{-3\tau/2}\ \
and\ \
|a_{i}(\tau)-\a_{2}e^{3\tau/2}|\leq C
\]
for $i=2,3$ and all $\tau\geq 0$.
\end{cor}

\textit{Proof}. Due to (\ref{eq:aitau}) and the fact that 
$\Sm=0$, the statements follow by arguments similar to those
given in the proof of Proposition \ref{prop:b8na}. $\hfill\Box$

\textit{Proof of Theorem \ref{thm:b8n}}.
Due to (\ref{eq:tp})
there is a constant $c_{\theta}>0$ such that 
\[
|\frac{1}{\theta}-c_{\theta}e^{3\tau/2}|\leq C.
\]
By (\ref{eq:ttau}) we conclude that 
\[
t(\tau)=2c_{\theta}e^{3\tau/2}+\delta(\tau)
\]
where 
\[
|\delta(\tau)|\leq C\tau
\]
for $\tau\geq 1$. The theorem follows from these observations and the 
previous corollary. $\hfill\Box$

\section{Approximations for general Bianchi VIII}\label{section:appr}

In this section we exclusively consider non-NUT Bianchi VIII solutions. 
The behaviour of a such solutions as $\tau\rightarrow \infty$
is in some sense oscillatory. The quantities that oscillate are $\Sm$ 
and $\Nt-\Nth$, and it will turn out that the frequency of the 
oscillations goes to infinity exponentially. Note that these
expressions go to zero as $\tau\rightarrow \infty$ by Theorem \ref{thm:b8}. 
On the other hand, $\Spp$ and $\No(\Nt+\Nth)$ do not converge to zero
and their derivatives are bounded. In order to analyze this situation,
we will proceed as follows. First we will approximate the oscillations
and try to write the quantities that oscillate as multiples of sine
and cosine. Then we use these approximations in the expressions for
the derivatives of the slowly varying quantities. Integrating over 
a period, we get an iteration. In order to be able to carry this out, we
will however need to know something concerning the variation of
different objects within a time interval corresponding to a period,
and we will have to spend some time writing down such estimates. 
Finally, we reformulate the iterations in the way most suitable to 
our purposes. 

Let 
\begin{equation}\label{eq:txtydef}
\tx=\Sm,\ \ty=\frac{\sqrt{3}}{2}(\Nt-\Nth).
\end{equation}
We have
\begin{equation}\label{eq:txpr}
\tx'=-3(\Nt+\Nth)\ty+\epsilon_{x},\
\epsilon_{x}=-(2-q)\Sm+3\No\ty.
\end{equation}
Furthermore,
\begin{equation}\label{eq:typr}
\ty'=3(\Nt+\Nth)\tx+\epsilon_{y},\
\epsilon_{y}=(q+2\Spp)\ty.
\end{equation}
Observe that by the constraint (\ref{eq:constraint}) $|\tx|,
|\ty|,|\epsilon_{x}|$ and $|\epsilon_{y}|$ are bounded by
numerical constants whereas $\Nt+\Nth\rightarrow \infty$ by
Theorem \ref{thm:b8}. 
Let $g=3(\Nt+\Nth)$,
\[
A=\left( \begin{array}{rr}
                0 & -g \\
                g & 0
            \end{array}
     \right)
\]
$\tbx={}^{t}(\tx, \ty) $ and
$\epsilon={}^{t}(\epsilon_{x}, \epsilon_{y})$ so that
$\tbx'=A\tbx+\epsilon$.
Let 
\begin{equation}\label{eq:xidef}
\xi(\tau)=\int_{\tau_{0}}^{\tau}g(s) d s+\phi_{0}
\end{equation}
for some $\phi_{0}$ and
$x_{1}(\tau)=\cos[\xi(\tau)],\
y_{1}(\tau)=\sin[\xi(\tau)].$
Then, if $\mathbf{x}_{1}={}^{t}(x_{1}, y_{1})$,
$\mathbf{x}_{1}'=A\mathbf{x}_{1}$. Define
\[
\Phi=\left( \begin{array}{rr}
                x_{1} & y_{1} \\
                -y_{1} & x_{1}
            \end{array}
     \right).
\]
Then $\Phi'=-A\Phi$ and $[A,\Phi]=0$. Let 
\begin{equation}\label{eq:rdef}
r(\tau)=[\tilde{x}^{2}(\tau)+\tilde{y}^{2}(\tau)]^{1/2} 
\end{equation}
and
\begin{equation}\label{eq:xdef}
\mathbf{x}(\tau)={}^{t}[x(\tau),y(\tau)]={}^{t}(r(\tau_{0})\cos[\xi(\tau)],
r(\tau_{0})\sin[\xi(\tau)])
\end{equation}
where  $\phi_{0}$ has been chosen so that $\mathbf{x}(\tau_{0})=
\tilde{\mathbf{x}}(\tau_{0})$. Since
$[\Phi (\mathbf{x}-\tilde{\mathbf{x}})]'=-\Phi\epsilon$
and $\Phi(\tau)\in SO(2)\ \forall \tau$ we have
\begin{equation}\label{eq:prel}
\|\tilde{\mathbf{x}}(\tau)-\mathbf{x}(\tau)\|\leq
|\int_{\tau_{0}}^{\tau}\|\epsilon(s)\| d s|. 
\end{equation}
We are mainly interested in intervals $[\tau_{1},\tau_{2}]$ such
that 
\begin{equation}\label{eq:twopi}
\xi(\tau_{2})-\xi(\tau_{1})=2\pi.
\end{equation}
Note that if $\tau_{\mathrm{min}}\in [\tau_{1},\tau_{2}]$ corresponds
to the minimum of $\Nt+\Nth$ in this interval, then
\[
\tau_{2}-\tau_{1}\leq\frac{2\pi}{3(\Nt+\Nth)(\tau_{\mathrm{min}})}.
\]
Furthermore, since $\Spp\rightarrow 1/2$ and $\Sm\rightarrow 0$, due to 
Theorem \ref{thm:b8}, there is for every $\epsilon>0$ a $T_{\epsilon}$
such that 
\[
\exp[(3/2-\epsilon)\tau]\leq
(\Nt+\Nth)(\tau)\leq\exp[(3/2+\epsilon)\tau]
\]
for all $\tau\geq T_{\epsilon}$. The corresponding bounds on 
$\tau_{2}-\tau_{1}$ will be used without further notice in the following.
Let us begin by considering the variation of $r$ in a time interval
$[\tau_{1},\tau_{2}]$ satisfying (\ref{eq:twopi}). 

\begin{lemma}\label{lemma:varr}
Let $[\tau_{1},\tau_{2}]$ be a time interval such that
(\ref{eq:twopi}) is fulfilled. If $\tau_{1}$ is big enough, depending
on the initial data, then if $\tau_{\mathrm{min}}$
corresponds to the minimum of $r$ in this interval,
\begin{equation}\label{eq:rvar}
|r(\tau_{a})-r(\tau_{b})|\leq
 2r(\tau_{\mathrm{min}})|\tau_{2}-\tau_{1}|
\end{equation}
for any $\tau_{a},\tau_{b}\in [\tau_{1},\tau_{2}]$. Furthermore,
\begin{equation}\label{eq:xvar}
\|\tilde{\mathbf{x}}(\tau)-\mathbf{x}(\tau)\|\leq
2r(\tau_{\mathrm{min}})|\tau_{2}-\tau_{1}|,
\end{equation}
for $\tau\in [\tau_{1},\tau_{2}]$
where the $\tau_{0}\in [\tau_{1},\tau_{2}]$ needed to define
$\mathbf{x}$ is arbitrary. 
\end{lemma}

\textit{Proof}. Let $\tau_{\mathrm{max}}$ correspond to a maximum of $r$ in
the interval. Since $\Spp\rightarrow 1/2$, $\Sm\rightarrow 0$ and
$\No\rightarrow 0$, we get
\begin{equation}\label{eq:epest}
\|\epsilon(s)\|^{2}\leq \frac{10}{4}r^{2}(s)\leq
\frac{10}{4}r^{2}(\tau_{\mathrm{max}})
\end{equation}
for $s\in [\tau_{1},\tau_{2}]$, assuming $\tau_{1}$ big enough.
Thus, by (\ref{eq:prel}), we have
\[
|r(\tau)-r(\tau_{\mathrm{max}})|\leq 
\left(\frac{10}{4}\right)^{1/2}r(\tau_{0})|\tau-\tau_{\mathrm{max}}|
\]
assuming we have chosen $\tau_{0}=\tau_{\mathrm{max}}$ when defining
of $\mathbf{x}$. Since $|\tau_{2}-\tau_{1}|$ can be assumed to be 
arbitrarily small if $\tau_{1}$ is big enough, (\ref{eq:epest})
implies
\[
\|\epsilon(s)\|\leq 2r(\tau_{\mathrm{min}}).
\]
The estimates (\ref{eq:rvar}) and (\ref{eq:xvar}) follow. $\hfill\Box$

The next lemma collects some technical estimates. We present them
here in order not to interrupt the flow of later proofs.

\begin{lemma}\label{lemma:tnest}
Consider a Bianchi VIII solution of
(\ref{eq:whsu})-(\ref{eq:constraint}) and let $C$ be a constant. Then
there is a $T$ depending on $C$ and the initial values such that if 
$[\tau_{1},\tau_{2}]$ is a time interval with $\tau_{1}\geq T$
and  
\[
|\tau_{2}-\tau_{1}|\leq \frac{C}{(\Nt+\Nth)(\tau_{3})}
\]
for some $\tau_{3}\in [\tau_{1},\tau_{2}]$, then
\begin{eqnarray}
|\Spp(t_{1})-\Spp(t_{2})|  \leq  \frac{C_{1}}{(\Nt+\Nth)(t_{3})}
\label{eq:spvar}\\
|[\No(\Nt+\Nth)](t_{1})-[\No(\Nt+\Nth)](t_{2})| 
 \leq  \frac{C_{2}}{(\Nt+\Nth)(t_{3})}\label{eq:nvar}\\
|1-\frac{(\Nt+\Nth)(t_{1})}{(\Nt+\Nth)(t_{2})}|  \leq 
\frac{C_{3}}{(\Nt+\Nth)(t_{3})}\label{eq:n23var}
\end{eqnarray}
for arbitrary $t_{1},t_{2},t_{3}\in [\tau_{1},\tau_{2}]$ where 
$C_{1}$, $C_{2}$ and $C_{3}$ are  constants only depending on $C$.
\end{lemma}
\textit{Proof}. We can assume $T$ to be 
great enough that $\Delta\tau=\tau_{2}-\tau_{1}\leq 1$. The inequality
\[
|\frac{(\Nt+\Nth)'}{\Nt+\Nth}|\leq 8
\]
follows from (\ref{eq:whsu})-(\ref{eq:constraint}). Thus
\[
|\frac{(\Nt+\Nth)(t_{1})}{(\Nt+\Nth)(t_{2})}|\leq
  e^{8(\tau_{2}-\tau_{1})}
\]
so that if $(\Nt+\Nth)(t_{1})\geq (\Nt+\Nth)(t_{2})$,
\begin{equation}\label{eq:n23varp}
|1-\frac{(\Nt+\Nth)(t_{1})}{(\Nt+\Nth)(t_{2})}|\leq
| e^{8\Delta\tau}-1|\leq  e^{8}\Delta\tau
\end{equation}
since $\Delta\tau\leq 1$. Multiplying this inequality by
$(\Nt+\Nth)(t_{2})/(\Nt+\Nth)(t_{1})\leq 1$ we conclude that the
assumption  $(\Nt+\Nth)(t_{1})\geq (\Nt+\Nth)(t_{2})$ is
not essential. There is thus 
a constant $c<\infty$ such that
\[
\frac{(\Nt+\Nth)(t_{1})}{(\Nt+\Nth)(t_{2})}\leq c
\]
for any $t_{1},t_{2}\in [\tau_{1},\tau_{2}]$. Consequently
\begin{equation}\label{eq:detest}
\Delta\tau\leq \frac{C}{(\Nt+\Nth)(\tau_{3})}\leq
\frac{cC}{(\Nt+\Nth)(t_{3})}.
\end{equation}
By (\ref{eq:whsu})-(\ref{eq:constraint}) 
$|\Spp'|$ and $|[\No(\Nt+\Nth)]'|$ are bounded by  
constants independent of Bianchi VIII solution. Equations
(\ref{eq:spvar}) and (\ref{eq:nvar}) follow from (\ref{eq:detest}).
Equation (\ref{eq:n23var}) follows from (\ref{eq:n23varp}) and
(\ref{eq:detest}).
$\hfill\Box$

\begin{lemma}\label{lemma:int}
Let $[\tau_{1},\tau_{2}]$ be a time interval such that
(\ref{eq:twopi}) holds. Then, if $\tau_{1}$ is big enough,
there is a constant $C>0$ such that
\begin{equation}\label{eq:sinkv}
\int_{\tau_{1}}^{\tau_{2}}\sin^{2} \xi(\tau)d\tau=
\frac{1}{2}(\tau_{2}-\tau_{1})+\eta_{1}
\end{equation}
and
\begin{equation}\label{eq:sinkub}
\int_{\tau_{1}}^{\tau_{2}}\sin^{4} \xi(\tau)d\tau=
\frac{3}{8}(\tau_{2}-\tau_{1})+\eta_{2}
\end{equation}
where $|\eta_{i}|\leq C/(\Nt+\Nth)^{2}(\tau_{3})$
for any $\tau_{3}\in [\tau_{1},\tau_{2}]$.
\end{lemma}
\textit{Proof}. We have
\[
\int_{\tau_{1}}^{\tau_{2}}\sin^{2} \xi(\tau)d\tau=
\int_{\xi(\tau_{1})}^{\xi(\tau_{2})}\sin^{2}\xi\frac{d\xi}{g}=
\frac{\pi}{g(\tau_{1})}+\delta_{1}
\]
where $\delta_{1}$ satisfies an estimate of the same form as $\eta_{1}$ by
(\ref{eq:n23var}). However
\[
\tau_{2}-\tau_{1}=\int_{\xi(\tau_{1})}^{\xi(\tau_{2})}\frac{d\xi}{g}
=\frac{2\pi}{g(\tau_{1})}+\delta_{2},
\]
where $\delta_{2}$ is of the same type as $\delta_{1}$. Equation 
(\ref{eq:sinkv}) follows, and the 
argument to prove (\ref{eq:sinkub}) is similar. $\hfill\Box$

Consider 
\[
h=r^2=\tx^2+\ty^2.
\]

\begin{lemma}\label{lemma:dhprel}
Consider a Bianchi VIII solution to
(\ref{eq:whsu})-(\ref{eq:constraint}). There is a constant $C>0$, such
that if $\tau_{1}$ is big enough and $[\tau_{1},\tau_{2}]$ is an
interval such that (\ref{eq:twopi}) is fulfilled, then
\[
h_{2}-h_{1}=[4(z_{1}-\frac{1}{2})(1+z_{1})+2h_{1}+\beta]
h_{1}\cdot (\tau_{2}-\tau_{1})
\]
where $z_{1}=\Spp(\tau_{1})$, $h_{i}=h(\tau_{i})$, and
$|\beta|\leq C/(\Nt+\Nth)(\tau_{1})$.
\end{lemma}

\textit{Proof}. We have
\[
\frac{dh}{d\tau}=2\tx\tx'+2\ty\ty'=2\tx[-3(\Nt+\Nth)\ty+\epsilon_{x}]
+2\ty[3(\Nt+\Nth)\tx+\epsilon_{y}]=
\]
\[
=2\tx\epsilon_{x}+2\ty\epsilon_{y}=-2(2-q)\tx^2+6\No\tx\ty+
2(q+2\Spp)\ty^2=
\]
\[
=-4(1-\Spp^2)\tx^2+4\tx^4+6\No\tx\ty+
4\Spp(1+\Spp)\ty^2+4\tx^2\ty^2.
\]
Consider an interval $[\tau_{1},\tau_{2}]$ such that (\ref{eq:twopi})
is fulfilled and let $z_{1}=\Spp(\tau_{1})$. In such an interval, we
have, using Lemma \ref{lemma:varr}, \ref{lemma:tnest} and the fact
that $|\No|\leq C/(\Nt+\Nth)$,
\[
\frac{dh}{d\tau}=-4(1-z_{1}^{2})x^2+4x^4+4z_{1}(1+z_{1})y^2+4x^2 y^2
+\delta
\]
where we have chosen $\tau_{0}=\tau_{1}$ in the definition of
$\mathbf{x}$ and $\delta$ satisfies
\[
|\delta|\leq C\frac{r^{2}(\tau_{1})}{(\Nt+\Nth)(\tau_{1})}
\]
where $C$ is a numerical constant for $\tau_{1}$ big enough. Applying
Lemma \ref{lemma:int}, we conclude that
\[
h_{2}-h_{1}=[-2(1-z_{1}^{2})+\frac{3}{2}h_{1}+2z_{1}(1+z_{1})+
\frac{1}{2}h_{1}]h_{1}(\tau_{2}-\tau_{1})+\eta
\]
where 
$|\eta|\leq Ch_{1}\cdot(\tau_{2}-\tau_{1})/(\Nt+\Nth)(\tau_{1})$
and $h_{i}=h(\tau_{i})$. The lemma follows. $\hfill\Box$

It will be convenient to reformulate this expression. 
Observe that, by the constraint,
\[
h+\frac{3}{4}\No^2+\frac{3}{2}(w-\frac{1}{2})=(\frac{1}{2}-z)
(\frac{1}{2}+z)=-(z-\frac{1}{2})-(z-\frac{1}{2})^{2}
\]
or
\begin{equation}\label{eq:conrel}
h+\frac{3}{4}\No^2+\frac{3}{2}v=-u-u^2,
\end{equation}
where
\[
z=\Spp,\ \ u=z-\frac{1}{2},\ \ w=-\No(\Nt+\Nth) \ \
\mathrm{and}\ \ \ v=w-\frac{1}{2}.
\]
\begin{lemma}
Consider a Bianchi VIII solution to
(\ref{eq:whsu})-(\ref{eq:constraint}). There is a constant $C>0$, such
that if $\tau_{1}$ is big enough and $[\tau_{1},\tau_{2}]$ is an
interval such that (\ref{eq:twopi}) is fulfilled, then
\begin{equation}\label{eq:dh}
h_{2}-h_{1}=[-4h_{1}-9v_{1}-2u_{1}^{2}+\delta]h_{1}(\tau_{2}-\tau_{1})
\end{equation}
where $u_{1}=u(\tau_{1})$, $v_{1}=v(\tau_{1})$, $h_{i}=h(\tau_{i})$, and
$|\delta|\leq C/(\Nt+\Nth)(\tau_{1})$.
\end{lemma}

\textit{Proof}. By (\ref{eq:conrel}) we conclude that 
\[
4(z-\frac{1}{2})(1+z)+2h=4u^{2}+6u+2h=
4u^{2}-6h-\frac{9}{2}\No^2-9v
-6u^{2}+2h=
\]
\[
=-4h-9v-2u^{2}-\frac{9}{2}\No^2.
\]
The lemma follows by applying Lemma \ref{lemma:dhprel}. $\hfill\Box$

Observe that, since $u$ converges to zero,
(\ref{eq:conrel}) implies 
\begin{equation}\label{eq:ub}
u^2\leq C_{1}h^{2}+C_{2}v^{2}
+C_{3}\No^4,
\end{equation}
so that the first two terms in the first factor on the right hand side
of (\ref{eq:dh}) are the important ones.

\begin{lemma}
Consider a Bianchi VIII solution to
(\ref{eq:whsu})-(\ref{eq:constraint}). There is a constant $C>0$, such
that if $\tau_{1}$ is big enough and $[\tau_{1},\tau_{2}]$ is an
interval such that (\ref{eq:twopi}) is fulfilled, then
\begin{equation}\label{eq:dv}
v_{2}-v_{1}=[h_{1}^{2}-\frac{3}{2}v_{1}+\psi(h_{1},u_{1},v_{1})+\delta]
(\tau_{2}-\tau_{1})
\end{equation}
where $|\delta|\leq C/(\Nt+\Nth)(\tau_{1})$.
Here
\begin{equation}\label{eq:psidef}
\psi(h,u,v)=[2h^2-3v+\phi(h,u,v)]v+\frac{1}{2}\phi(h,u,v),
\end{equation}
\begin{equation}\label{eq:phidef}
\phi(h,u,v)=2[\frac{9}{4}v^2+u^4+2hu^2+3hv+3vu^2],
\end{equation}
and we use the standard notation, i.e. $v_{i}=v(\tau_{i})$ etc.
\end{lemma}

\textit{Proof}. Consider
\[
w'=-[\No(\Nt+\Nth)]'=(2q-2\Spp)w-2\sqrt{3}\Sm\No(\Nt-\Nth).
\]
With the standard notations, we have
\[
w'=4z_{1}(z_{1}-\frac{1}{2})w_{1}+4x^2 w_{1}+\delta
\]
in $[\tau_{1},\tau_{2}]$, where
$|\delta|\leq C/(\Nt+\Nth)(\tau_{1})$. Consequently
\[
w_{2}-w_{1}=[4z_{1}(z_{1}-\frac{1}{2})+2h_{1}]w_{1}(\tau_{2}-\tau_{1})
+\delta_{1}
\]
where $|\delta_{1}|\leq C(\tau_{2}-\tau_{1})/(\Nt+\Nth)(\tau_{1})$.
Let us reformulate, using (\ref{eq:conrel}),
\[
4z(z-\frac{1}{2})+2h=2u+4u^2+2h
=-2u^2-2h-\frac{3}{2}\No^2-3v+
4u^2+2h=
\]
\[
=2u^2-3v-\frac{3}{2}\No^2=
2h^2-3v+\phi(h,u,v)+\delta_{2},
\]
where $\phi(h,u,v)$ is defined by (\ref{eq:phidef}),
so that in particular,
\[
|\phi(h,u,v)|\leq C_{1}v^2+C_{2}u^{4}+C_{3}hu^2+C_{4}h|v|
\]
Furthermore
$|\delta_{2}|\leq C(\tau_{2}-\tau_{1})/(\Nt+\Nth)(\tau_{1})$.
Thus
\[
v_{2}-v_{1}=[2h_{1}^{2}-3v_{1}+\phi(h_{1},u_{1},v_{1})+\delta_{2}]w_{1}
(\tau_{2}-\tau_{1})+\delta_{1}.
\]
Letting $\psi$ be defined by (\ref{eq:psidef}),
we get the conclusion of the lemma. $\hfill\Box$

It is nice to have an iteration, but it is nicer to have integral
expressions.
\begin{lemma}
Consider a Bianchi VIII solution to
(\ref{eq:whsu})-(\ref{eq:constraint}). There is a constant $C>0$ such
that if $\tau_{a}$ is big enough and
$[\tau_{a},\tau_{b}]$ is an interval
such that $\xi(\tau_{b})-\xi(\tau_{a})$ is an integer
multiple of $2\pi$, then
\begin{equation}\label{eq:dhint}
\frac{h(\tau_{b})}{h(\tau_{a})}=
\exp[-\int_{\tau_{a}}^{\tau_{b}}(4h+9v+2u^2)d\tau+\delta_{5}],
\end{equation}
where
\begin{equation}\label{eq:dhintest}
|\delta_{5}|\leq C\int_{\tau_{a}}^{\tau_{b}}e^{-s}ds.
\end{equation}
\end{lemma}
\textit{Proof}.
Consider (\ref{eq:dh}). Observe that
\[
|h_{1}\cdot (\tau_{2}-\tau_{1})-\int_{\tau_{1}}^{\tau_{2}}h(s)ds|\leq
\frac{C(\tau_{2}-\tau_{1})}{(\Nt+\Nth)(\tau_{1})}
\]
by (\ref{eq:rvar}), and similarly for the other expressions by
(\ref{eq:spvar}) and (\ref{eq:nvar}), so that 
\[
\frac{h_{2}}{h_{1}}-1=-\int_{\tau_{1}}^{\tau_{2}}(4h+9v+2u^2)d\tau+
\delta_{3}
\]
where
$|\delta_{3}|\leq C(\tau_{2}-\tau_{1})/(\Nt+\Nth)(\tau_{1})$.
Thus
\[
\frac{h_{2}}{h_{1}}=\exp[-\int_{\tau_{1}}^{\tau_{2}}(4h+9v+2u^2)d\tau+
\delta_{4}]
\]
where $\delta_{4}$ satisfies an estimate similar to that of $\delta_{3}$.
The reason is the following. Let
\[
a=-\int_{\tau_{1}}^{\tau_{2}}(4h+9v+2u^2)d\tau+
\delta_{3}.
\]
We have
\[
\frac{h_{2}}{h_{1}}=\exp[\ln (1+a)],
\]
so that we need to estimate $\ln(1+a)-a$. However,
\[
|\ln(1+a)-a|\leq a^{2}
\]
if $|a|\leq 1/2$. This criterion is fulfilled for $\tau_{1}$ big
enough, and the desired conclusion follows, since
\[
a^{2}\leq \frac{C(\tau_{2}-\tau_{1})}{(\Nt+\Nth)(\tau_{1})}.
\]
Observe furthermore that for $\tau_{1}$ big enough
\[
|\delta_{4}|\leq C\int_{\tau_{1}}^{\tau_{2}}e^{-s}ds,
\]
since $\Spp\rightarrow 1/2$ and $\Sm\rightarrow 0$ due to Theorem
\ref{thm:b8}. The lemma follows. $\hfill\Box$

Observe that (\ref{eq:dhint}) implies that $v$ and $h$ cannot both
be $L^{1}([\tau_{0},\infty))$, using the bound (\ref{eq:ub}).
Similarly, we get
\begin{equation}\label{eq:dvint}
v(\tau_{b})-v(\tau_{a})=\int_{\tau_{a}}^{\tau_{b}}
[h^{2}-\frac{3}{2}v+\psi(h,u,v)]ds+\delta_{6}
\end{equation}
where
\begin{equation}\label{eq:dvintest}
|\delta_{6}|\leq C\int_{\tau_{a}}^{\tau_{b}}e^{-s}ds.
\end{equation}

\section{Asymptotics of general Bianchi VIII}\label{section:b8as}

In the following we wish to use the tools we developed in the previous
section in order to analyze the details of the asymptotics of $h,u,v$.
Note that we already know that $(h,u,v)$ converges to zero by Theorem
\ref{thm:b8}. Let us start by bounding $v$ from below. 
\begin{lemma}
Consider a Bianchi VIII solution to
(\ref{eq:whsu})-(\ref{eq:constraint}). Let $0<\a<1$. Then
there is a $T_{\a}$ such that 
\begin{equation}\label{eq:vbb}
v(\tau)\geq -e^{-\a\tau}
\end{equation}
for all $\tau\geq T_{\a}$.
\end{lemma}

\textit{Proof}.
Consider (\ref{eq:dvint}). If $v$ is negative and $\tau$ is big
enough, then
\[
h^2-\frac{3}{2}v+\psi(h,u,v)\geq -v.
\]
Consider now the situation where $[\tau_{a},\tau_{b}]$ corresponds to
a single multiple of $2\pi$ and $\tau_{a}$ is big enough. Let 
$0<\a<\beta<1$, and assume that 
\[
|v(\tau_{a})|\geq e^{-\beta\tau_{a}}.
\]
Observe that under these assumptions, $v$ cannot change sign in 
$[\tau_{a},\tau_{b}]$ if $\tau_{a}$ is big enough, since the maximum 
variation of $v$ in a time interval corresponding to $2\pi$ is bounded 
by $C/(\Nt+\Nth)(\tau_{a})$. Since
\[
|\frac{v(\tau)-v(\tau_{a})}{v(\tau_{a})}|\leq Ce^{-(1-\beta)\tau_{a}}
\]
under these circumstances, assuming $\tau_{a}$ big enough, we
have
\[
\frac{v(\tau_{b})}{v(\tau_{a})}\leq 1-\int_{\tau_{a}}^{\tau_{b}}
\frac{v(\tau)}{v(\tau_{a})}d\tau+\frac{\delta_{6}}{v(\tau_{a})}
\leq 1-(\tau_{b}-\tau_{a})+
\delta_{7}
\]
where
\[
|\delta_{7}|\leq C\int_{\tau_{a}}^{\tau_{b}}e^{-(1-\beta)\tau}d\tau.
\]
Thus
\[
\frac{v(\tau_{b})}{v(\tau_{a})}\leq
\exp[-(\tau_{b}-\tau_{a})+
\delta_{8}]
\]
where $\delta_{8}$ satisfies an estimate similar to that of $\delta_{7}$.
Observe that we at this point can drop the restriction that 
$[\tau_{a},\tau_{b}]$ correspond to $2\pi$. It can be any
interval corresponding to an integer multiple of $2\pi$, as
long as $v(\tau)\leq -e^{-\beta\tau}$ in the interval.
If we assume that $v\leq-\exp(-\beta\tau)$ for all $\tau\geq T$,
we get a contradiction. At late enough times, once $v$ has
satisfied the inequality $v\geq-\exp(-\beta\tau)$, it cannot 
at a later point in time fulfill $v\leq-\exp(-\a\tau)$.
The lemma follows.  $\hfill\Box$

It will be convenient to have the following rough estimate.
\begin{lemma}\label{lemma:hbr}
Consider a Bianchi VIII solution to
(\ref{eq:whsu})-(\ref{eq:constraint}). For every $\a>0$,
there is a $T$ such that $\tau_{1},\tau_{2},\tau\geq T$ implies
\[
h(\tau)\geq e^{-\a\tau}
\]
and, assuming $\tau_{2}\geq\tau_{1}$ corresponds to an integral
multiple of $2\pi$,
\[
h(\tau_{2})\geq h(\tau_{1})\exp[-\a(\tau_{2}-\tau_{1})].
\]
\end{lemma}
\textit{Proof}. The lemma follows from the fact that the integrand
appearing in (\ref{eq:dhint}) tends to zero. $\hfill\Box$

\begin{lemma}\label{lemma:vb}
Consider a Bianchi VIII solution to
(\ref{eq:whsu})-(\ref{eq:constraint}). Then there is a $T$ such
that 
\[
|v(\tau)|\leq 2h^{2}(\tau)
\]
for all $\tau\geq T$.
\end{lemma}

\textit{Proof}. Let us first prove that for every $T$, there is 
$\tau\geq T$ such that $v(\tau)\leq\frac{10}{9} h^2(\tau)$.
Assume that there is a $T$ such that 
\begin{equation}\label{eq:a1}
v(\tau)\geq\frac{10}{9} h^2(\tau)
\end{equation}
for all $\tau\geq T$. 
Consider (\ref{eq:dvint}). If assumption (\ref{eq:a1}) is satisfied
and $\tau_{a}$ is big enough, we conclude that, 
\[
v(\tau_{b})-v(\tau_{a})\leq -\frac{1}{2}\int_{\tau_{a}}^{\tau_{b}}
vds
\]
where $[\tau_{a},\tau_{b}]$ corresponds to an integral multiple of
$2\pi$, and we have used (\ref{eq:dvint}) and Lemma \ref{lemma:hbr}.
Consider now the situation where $[\tau_{a},\tau_{b}]$ corresponds to
a single multiple of $2\pi$. Observe that
\[
|v(\tau)|\geq e^{-\tau/2}
\]
if $\tau$ big enough, using (\ref{eq:a1}) and Lemma
\ref{lemma:hbr}.
Since
\[
|\frac{v(\tau)-v(\tau_{a})}{v(\tau_{a})}|\leq Ce^{-\tau_{a}/2}
\]
for $\tau\in [\tau_{a},\tau_{b}]$
under these circumstances, assuming $\tau_{a}$ big enough, we
have
\[
\frac{v(\tau_{b})}{v(\tau_{a})}\leq 1-\frac{1}{2}(\tau_{b}-\tau_{a})+
\delta_{11}
\]
where
\[
|\delta_{11}|\leq C\int_{\tau_{a}}^{\tau_{b}}e^{-\tau/2}d\tau.
\]
Thus
\[
\frac{v(\tau_{b})}{v(\tau_{a})}\leq
\exp[-\frac{1}{2}(\tau_{b}-\tau_{a})+
\delta_{12}]
\]
where $\delta_{12}$ satisfies an estimate similar to that of 
$\delta_{11}$. Observe that we at this point can drop the restriction
that $[\tau_{a},\tau_{b}]$ correspond to $2\pi$. It can correspond to
any integer multiple of $2\pi$. Combining (\ref{eq:a1}) with
this estimate, we get a result incompatible with Lemma
\ref{lemma:hbr}. For every $T$, there must thus be a $\tau\geq T$
such that 
\[
v(\tau)\leq\frac{10}{9} h^2(\tau).
\]
Assume now that for every $T$, there is a $\tau\geq T$ such that
\[
v(\tau)\geq 2h^{2}(\tau).
\]
By the above, we can then find intervals $[\tau_{1},\tau_{2}]$ with
$\tau_{1}$ arbitrarily great,
\[
\frac{10}{9} h^2(\tau)\leq v(\tau)\leq 2h^{2}(\tau)
\]
for $\tau\in[\tau_{1},\tau_{2}]$ and 
\[
v(\tau_{1})=\frac{10}{9} h^2(\tau_{1})\ \ \ \mathrm{and}\ \ \
v(\tau_{2})=2h^{2}(\tau_{2}).
\]
Observe that the interval $[\tau_{1},\tau_{2}]$ will contain a
subinterval corresponding to an arbitrarily large number of 
multiples of $2\pi$ by choosing $\tau_{1}$ big enough. This is
due to Lemma \ref{lemma:hbr} and the fact that $v$ does not vary
more than roughly speaking $\exp[-3\tau/2]$ in a time interval 
corresponding to $2\pi$. Let $\tau_{2-}$ be such that 
$[\tau_{1},\tau_{2-}]$ corresponds to an integer multiple of 
$2\pi$, but $[\tau_{2-},\tau_{2}]$ corresponds to less than an
integer multiple of $2\pi$. For any $\a>0$, we can choose
an interval $[\tau_{1},\tau_{2}]$ as above such that 
Lemma \ref{lemma:hbr} yields
\[
h^{2}(\tau_{2-})\geq h^{2}(\tau_{1})\exp[-\a(\tau_{2-}-\tau_{1})].
\]
The above arguments yield
\[
v(\tau_{2-})\leq \frac{10}{9}h^{2}(\tau_{1})
\exp[-\frac{1}{2}(\tau_{2-}-\tau_{1})+\delta_{12}]
\]
with $\delta_{12}$ satisfying an estimate as above. Consequently
\[
v(\tau_{2-})\leq \frac{10}{9}h^{2}(\tau_{2-})
\]
if $\tau_{1}$ is big enough. Since the interval $[\tau_{2-},\tau_{2}]$
is to short to remedy this, the lemma follows, at least without the
absolute value sign. If $v$ is negative, we use (\ref{eq:vbb}) and
Lemma \ref{lemma:hbr}. $\hfill\Box$

\begin{lemma}
Consider a Bianchi VIII solution to
(\ref{eq:whsu})-(\ref{eq:constraint}) and let $0<\a_{1}<1<\a_{2}$ be
arbitrary numbers such that $\a_{1}\a_{2}=1$. Then, if $\tau_{a}$ is
big enough,
\[
\frac{\a_{1}h(\tau_{a})}{[1+4h(\tau_{a})(\tau-\tau_{a})]^{\a_{2}/\a_{1}}}\leq
h(\tau)\leq
\frac{\a_{2}h(\tau_{a})}{[1+4h(\tau_{a})(\tau-\tau_{a})]^{\a_{1}/\a_{2}}}
\]
for all $\tau\geq\tau_{a}$.
\end{lemma}

\textit{Proof}. Given $\a_{1}<1<\a_{2}$ such that 
$\a_{1}\a_{2}=1$, there is a $T$ such that if 
$\tau_{a}\geq T$ and $\tau\in (\tau_{a}-\epsilon_{a},\infty)$,
where $\epsilon_{a}>0$ depends on $\tau_{a}$, then
\begin{equation}\label{eq:ruhb}
\a_{1}\exp[-4\a_{2}\int_{\tau_{a}}^{\tau}h(s)ds]\leq
\frac{h(\tau)}{h(\tau_{a})}\leq
\a_{2}\exp[-4\a_{1}\int_{\tau_{a}}^{\tau}h(s)ds].
\end{equation}
Note first of all that it holds if $\tau\geq \tau_{a}$ is
such that $[\tau_{a},\tau]$ corresponds to an integer multiple
of $2\pi$, due to (\ref{eq:dhint}), Lemma \ref{lemma:vb},
(\ref{eq:ub}) and Lemma \ref{lemma:hbr}. However, moving around in
a time interval that corresponds to less than a multiple of 
$2\pi$ doesn't change any of the constituents of the equation all
that much. Let
\[
g(\tau)=4\int_{\tau_{a}}^{\tau}h(s)ds.
\]
By (\ref{eq:ruhb}), we have
\[
\a_{1}\exp[-\a_{2}g(\tau)]\leq
\frac{\dot{g}(\tau)}{\dot{g}(\tau_{a})}\leq
\a_{2}\exp[-\a_{1}g(\tau)].
\]
Integrating the left inequality, we conclude that
\[
1+\dot{g}(\tau_{a})(\tau-\tau_{a})\leq \exp[\a_{2}g(\tau)],
\]
and by integrating the right, we get
\[
\exp[\a_{1}g(\tau)]\leq 1+\dot{g}(\tau_{a})(\tau-\tau_{a}).
\]
Inserting these inequalities in (\ref{eq:ruhb}) yields the
conclusion of the lemma. $\hfill\Box$
\begin{cor}\label{cor:prel}
Consider a Bianchi VIII solution to
(\ref{eq:whsu})-(\ref{eq:constraint}) and let $0<\a_{1}<1<\a_{2}$ be
arbitrary numbers. Then, if $\tau$ is great enough,
\[
\tau^{-\a_{2}}\leq h(\tau)\leq\tau^{-\a_{1}}.
\]
\end{cor}

\begin{prop}\label{prop:has1}
Consider a Bianchi VIII solution to
(\ref{eq:whsu})-(\ref{eq:constraint}). There is a $T$ and a constant
$C>0$, such that for all $\tau\geq T$, 
\[
|h(\tau)-\frac{1}{4\tau}|\leq\frac{C\ln \tau}{\tau^{2}}.
\]
\end{prop}

\textit{Proof}. By the proof of Lemma \ref{lemma:dhprel}, 
\[
\frac{dh}{d\tau}=-4(1-\Spp^2)\tx^2+4\tx^4+6\No\tx\ty+
4\Spp(1+\Spp)\ty^2+4\tx^2\ty^2,
\]
so that
\[
\frac{1}{h^2}\frac{dh}{d\tau}=-4(1-\Spp^2)\frac{\tx^2}{h^2}+
4\frac{\tx^4}{h^2}+6\No\frac{\tx\ty}{h^2}+
4\Spp(1+\Spp)\frac{\ty^2}{h^2}+4\frac{\tx^2\ty^2}{h^2}.
\]
Consider a time interval $[\tau_{1},\tau_{2}]$ such that 
(\ref{eq:twopi}) is satisfied. Observe that
\[
|\frac{1}{h^2(\tau)}-\frac{1}{h^2(\tau_{1})}|=
\frac{[h(\tau)+h(\tau_{1})]|h(\tau)-h(\tau_{1})|}
{h^2(\tau)h^2(\tau_{1})}.
\]
Since 
\[
|h(\tau)-h(\tau_{1})|\leq Ch(\tau_{1})|\tau_{2}-\tau_{1}|
\]
by (\ref{eq:rvar}), and Corollary \ref{cor:prel} holds,
we conclude that 
\[
|\frac{1}{h^2(\tau)}-\frac{1}{h^2(\tau_{1})}|\leq Ce^{-\tau_{1}}
\]
for $\tau\in[\tau_{1},\tau_{2}]$, if $\tau_{1}$ is big enough.
Due to Corollary \ref{cor:prel}, Lemma \ref{lemma:tnest} and
(\ref{eq:xvar}), we also conclude that 
\[
\frac{1}{h^2}\frac{dh}{d\tau}=-4(1-z_{1}^2)\frac{x^2}{h_{1}^2}+
4\frac{x^4}{h_{1}^2}+
4z_{1}(1+z_{1})\frac{y^2}{h_{1}^2}+4\frac{x^2
y^2}{h_{1}^2}+\delta_{13},
\]
where $|\delta_{13}|\leq C\exp(-\tau_{1})$. By Lemma \ref{lemma:int}
we conclude that 
\[
-\frac{1}{h(\tau_{2})}+\frac{1}{h(\tau_{1})}=
[-2(1-z_{1}^2)\frac{1}{h_{1}}+
\frac{3}{2}+
2z_{1}(1+z_{1})\frac{1}{h_{1}}+\frac{1}{2}](\tau_{2}-\tau_{1})+
\delta_{14}
\]
where 
\[
|\delta_{14}|\leq C\int_{\tau_{1}}^{\tau_{2}}e^{-\tau}d\tau.
\]
Let us now reformulate 
\[
-2(1-z_{1}^2)+2z_{1}(1+z_{1})=4(z_{1}-\frac{1}{2})(z_{1}+1)=
6u_{1}+4u_{1}^2.
\]
Using (\ref{eq:conrel}), we get
\[
6u_{1}+4u_{1}^2=-6u_{1}^{2}-6h_{1}-\frac{9}{2}\No^{2}(\tau_{1})
-9v_{1}+4u_{1}^{2}=[-6+\delta_{15}]h_{1}
\]
where $|\delta_{15}|\leq Ch_{1}$,
due to Lemma \ref{lemma:vb}, Corollary \ref{cor:prel} and
(\ref{eq:ub}). We conclude that 
\[
-\frac{1}{h(\tau_{2})}+\frac{1}{h(\tau_{1})}=
-4(\tau_{2}-\tau_{1})+\delta_{16}
\]
where
\[
|\delta_{16}|\leq C\int_{\tau_{1}}^{\tau_{2}}h(s)ds.
\]
At this point, we can drop the restriction that $[\tau_{1},\tau_{2}]$
correspond to one multiple of $2\pi$. The statement holds as long as
it corresponds to an integer multiple of $2\pi$. Fix $\tau_{1}$
big enough so that the above analysis holds, and define
\[
\zeta(\tau)=\frac{1}{h(\tau)}-4\tau.
\]
If $[\tau_{1},\tau_{2}]$ corresponds to an integral multiple of 
$2\pi$, then 
\[
\zeta(\tau_{2})=\frac{1}{h(\tau_{1})}-4\tau_{1}-\delta_{16}.
\]
Fix $\a>0$. Using Corollary \ref{cor:prel}, and assuming $\tau$ to be
big enough, we conclude that 
\[
|\zeta(\tau)|\leq C_{\a}\tau^{\a}.
\]
At this point, we see that the same inequality holds without
any restriction on $\tau$ except that it be big enough. Going through
the same argument using this improved knowledge concerning the 
asymptotic behaviour of $h$, we conclude that for $\tau$ great enough,
$|\zeta(\tau)|\leq C\ln \tau$. The proposition follows. $\hfill\Box$

\begin{prop}\label{prop:vas1}
Consider a Bianchi VIII solution to
(\ref{eq:whsu})-(\ref{eq:constraint}). There is a $T$ and a constant
$C>0$ such that 
\[
|v(\tau)-\frac{1}{24\tau^{2}}|\leq \frac{C\ln\tau}{\tau^{3}}
\]
for all $\tau\geq T$.
\end{prop}

\textit{Proof}. Let
\[
\chi=h^2-\frac{3}{2}v
\]
and consider an interval $[\tau_{1},\tau_{2}]$ such that
(\ref{eq:twopi}) is satisfied. Use (\ref{eq:dh}) and 
(\ref{eq:rvar}) in order to deduce 
\[
h_{2}^{2}-h^{2}_{1}=(h_{2}+h_{1})(h_{2}-h_{1})=
(h_{2}+h_{1})
[-4h_{1}-9v_{1}-2u_{1}^{2}+\delta]h_{1}(\tau_{2}-\tau_{1})=
\]
\[
=2h_{1}^{2}[-4h_{1}-9v_{1}-2u_{1}^{2}+\delta_{17}](\tau_{2}-\tau_{1}),
\]
where $|\delta_{17}|\leq C/(\Nt+\Nth)(\tau_{1})$.
Combining this with (\ref{eq:dv}), we get the conclusion
\[
\chi_{2}-\chi_{1}=[-\frac{3}{2}\chi_{1}-\frac{3}{2}
\psi(h_{1},u_{1},v_{1})-8h_{1}^{3}-18v_{1}h_{1}^{2}-
4u_{1}^{2}h_{1}^{2}+\delta_{18}](\tau_{2}-\tau_{1}),
\]
where $\delta_{18}$ satisfies a bound similar to that of 
$\delta_{17}$. Similarly to earlier arguments, we conclude that 
if $[\tau_{a},\tau_{b}]$ corresponds to an integer multiple of
$2\pi$ and $\tau_{a}$ is big enough, then 
\[
\chi(\tau_{b})-\chi(\tau_{a})=\int_{\tau_{a}}^{\tau_{b}}
[-\frac{3}{2}\chi-\frac{3}{2}
\psi(h,u,v)-8h^{3}-18vh^{2}-
4u^{2}h^{2}]d\tau+\delta_{19},
\]
where
\[
|\delta_{19}|\leq C\int_{\tau_{a}}^{\tau_{b}}e^{-\tau}d\tau.
\]
Using Lemma \ref{lemma:vb}, (\ref{eq:ub}), Proposition \ref{prop:has1}
and the definition of $\psi$, one deduces
\[
\chi(\tau_{b})-\chi(\tau_{a})=
-\frac{3}{2}\int_{\tau_{a}}^{\tau_{b}}\chi d\tau+\delta_{20},
\]
Where
\[
|\delta_{20}|\leq C_{1}\int_{\tau_{a}}^{\tau_{b}}\tau^{-3}d\tau.
\]
Assuming that there is a $T$ such that $\tau\geq T$ implies 
\[
|\chi(\tau)|\geq 2C_{1}\tau^{-3}
\]
yields the conclusion
\[
\chi(\tau_{b})-\chi(\tau_{a})\leq
-\int_{\tau_{a}}^{\tau_{b}}\chi d\tau,
\]
if $\chi$ is positive, and
\[
\chi(\tau_{b})-\chi(\tau_{a})\geq
-\int_{\tau_{a}}^{\tau_{b}}\chi d\tau,
\]
if $\chi$ is negative, assuming $\tau_{a}$ is big enough.
Arguments similar to ones given in the proof of Lemma \ref{lemma:vb} 
yield the conclusion that 
\[
|\chi(\tau)|\leq C\tau^{-3}
\]
if $\tau$ is big enough. The proposition follows. $\hfill\Box$

\begin{prop}
Consider a Bianchi VIII solution to
(\ref{eq:whsu})-(\ref{eq:constraint}). There are constants
$C>0$ and $c_{h,1}$ and a $T$ such that
\begin{equation}\label{eq:has2}
|h(\tau)-\frac{1}{4\tau}+\frac{\ln\tau}{8\tau^{2}}-
\frac{c_{h,1}}{\tau^{2}}|\leq\frac{C\ln^{2}\tau}{\tau^{3}}
\end{equation}
for all $\tau\geq T$.
\end{prop}

\textit{Proof}. Consider
\[
\th(\tau)=g(\tau)(\tau h(\tau)-\frac{1}{4})
\]
where
\[
g(\tau)=\exp(\int_{1}^{\tau}4h(s)ds-c_{1})
\]
and
\[
c_{1}=\int_{1}^{\infty}(4h(s)-\frac{1}{s})ds.
\]
That the integral defining $c_{1}$ converges follows from
Proposition \ref{prop:has1}. As a consequence, we have
\[
|\int_{1}^{\tau}(4h(s)-\frac{1}{s})ds-c_{1}|\leq
C\int_{\tau}^{\infty}\frac{\ln s}{s^{2}}ds\leq C\frac{\ln\tau}{\tau}.
\]
Thus
\[
|g(\tau)-\tau|=\tau
 |\exp[\int_{1}^{\tau}(4h(s)-\frac{1}{s})ds-c_{1}]-1|\leq
C\ln\tau
\]
for $\tau$ great enough.
Let $[\tau_{1},\tau_{2}]$ be an interval such that (\ref{eq:twopi})
is satisfied. With the usual notation, we have
\begin{equation}\label{eq:dhtprel}
\th_{2}-\th_{1}=(g_{2}-g_{1})(\tau_{2}h_{2}-\frac{1}{4})+
g_{1}(\tau_{2}h_{2}-\tau_{1}h_{1})=
\end{equation}
\[
=(g_{2}-g_{1})(\tau_{2}h_{2}-\tau_{1}h_{1})+
(g_{2}-g_{1})(\tau_{1}h_{1}-\frac{1}{4})+
g_{1}(\tau_{2}h_{2}-\tau_{1}h_{1}).
\] 
Compute
\[
g_{2}-g_{1}=g_{1}[\exp(\int_{\tau_{1}}^{\tau_{2}}4h(s)ds)-1]=
4g_{1}h_{1}(\tau_{2}-\tau_{1})+\delta_{20},
\]
where $\delta_{20}$ is of the order of magnitude
$(\tau_{2}-\tau_{1})^{2}$. 
Consider
\[
\tau_{2}h_{2}-\tau_{1}h_{1}=(\tau_{2}-\tau_{1})h_{2}+
\tau_{1}(h_{2}-h_{1})=
\]
\[
=(\tau_{2}-\tau_{1})(h_{2}-h_{1})+
(\tau_{2}-\tau_{1})h_{1}+\tau_{1}(h_{2}-h_{1}).
\]
Observe that, even if we multiply the first term with $g_{1}$, it
is still of the order of magnitude $(\tau_{2}-\tau_{1})^{2}$, using
(\ref{eq:dh}), and will for this reason be of no importance. By
(\ref{eq:dh}), we conclude
\[
(\tau_{2}-\tau_{1})h_{1}+\tau_{1}(h_{2}-h_{1})=
\]
\[
=[h_{1}-4\tau_{1}h_{1}^{2}-
9\tau_{1}v_{1}h_{1}-2\tau_{1}u_{1}^{2}h_{1}+\tau_{1}\delta h_{1}]
(\tau_{2}-\tau_{1})
\]
Consider the first term on the far right hand side of
(\ref{eq:dhtprel}). By the above it will be of the order of magnitude
$(\tau_{2}-\tau_{1})^{2}$. We conclude that 
\[
\th_{2}-\th_{1}=
\]
\[
=[4g_{1}h_{1}(\tau_{1}h_{1}-\frac{1}{4})+
g_{1}h_{1}-4\tau_{1}g_{1}h_{1}^{2}-
9\tau_{1}g_{1}v_{1}h_{1}-2\tau_{1}g_{1}u_{1}^{2}h_{1}](\tau_{2}-\tau_{1})
+\delta_{21}
\]
where
\[
|\delta_{21}|\leq C\int_{\tau_{1}}^{\tau_{2}}e^{-s}ds.
\]
In other words,
\[
\th_{2}-\th_{1}
=[
-9\tau_{1}g_{1}v_{1}h_{1}-2\tau_{1}g_{1}u_{1}^{2}h_{1}](\tau_{2}-\tau_{1})
+\delta_{21}.
\]
As in previous arguments, we conclude that 
\[
\th(\tau_{b})-\th(\tau_{a})=
\int_{\tau_{a}}^{\tau_{b}}[-9\tau gvh-2\tau gu^{2}h]ds +\delta_{22}
\]
where
\[
|\delta_{22}|\leq C\int_{\tau_{a}}^{\tau_{b}}e^{-s}ds
\]
and $[\tau_{a},\tau_{b}]$ corresponds to an integer multiple of 
$2\pi$. Observe that 
\[
-9\tau gvh-2\tau gu^{2}h=-\frac{9\tau^2}{24\tau^{2}\cdot 4\tau}-
\frac{2\tau^{2}}{64\tau^{3}}+O(\frac{\ln\tau}{\tau^{2}})=
-\frac{1}{8\tau}+O(\frac{\ln\tau}{\tau^{2}})
\]
In other words, we see that 
\[
\th(\tau)+\frac{1}{8}\ln\tau
\]
converges. Let $c_{h,1}$ denote the value to which it converges.
Then by the above estimates, we have
\[
|\th(\tau)+\frac{1}{8}\ln\tau-c_{h,1}|\leq C\frac{\ln\tau}{\tau}.
\]
In other words,
\begin{equation}\label{eq:has2prel}
|h(\tau)-\frac{1}{4\tau}+\frac{1}{8}\frac{\ln\tau}{\tau g(\tau)}-
\frac{c_{h,1}}{\tau g(\tau)}| \leq C\frac{\ln\tau}{\tau^{3}}.
\end{equation}
However, the above estimates imply
\[
|\frac{\ln\tau}{\tau g(\tau)}-
\frac{\ln\tau}{\tau^{2}}|\leq C\frac{\ln^{2}\tau}{\tau^{3}}
\]
and 
\[
|\frac{1}{\tau g(\tau)}-\frac{1}{\tau^{2}}|\leq
C\frac{\ln\tau}{\tau^{3}}
\]
so that the proposition follows. $\hfill\Box$

Observe that the estimate (\ref{eq:has2}) can be improved in the
following way. Using (\ref{eq:has2}), we conclude that 
\[
\int_{1}^{\tau}(4h(s)-\frac{1}{s})ds-c_{1}=
-\int_{\tau}^{\infty}(4h(s)-\frac{1}{s})ds=
\int_{\tau}^{\infty}[\frac{\ln s}{2 s^2}+O(\frac{1}{s^{2}})]ds.
\]
Consequently
\[
|\int_{1}^{\tau}(4h(s)-\frac{1}{s})ds-c_{1}-\frac{\ln\tau}{2\tau}|
\leq\frac{C}{\tau}.
\]
Thus
\[
|g(\tau)-\tau-\frac{1}{2}\ln\tau|=\tau
|\exp[\int_{1}^{\tau}(4h(s)-\frac{1}{s})ds-c_{1}-\frac{\ln\tau}{2\tau}]
\exp(\frac{\ln\tau}{2\tau})-1-\frac{\ln\tau}{2\tau}|.
\]
By the above
\[
|\exp[\int_{1}^{\tau}(4h(s)-\frac{1}{s})ds-c_{1}-\frac{\ln\tau}{2\tau}]-1|
\leq\frac{C}{\tau}.
\]
We also have
\[
|\exp(\frac{\ln\tau}{2\tau})-1-\frac{\ln\tau}{2\tau}|\leq
C\frac{\ln^{2}\tau}{\tau^{2}}.
\]
Adding up, we get 
\[
|g(\tau)-\tau-\frac{1}{2}\ln\tau|\leq C.
\]
Thus
\[
|\frac{1}{g(\tau)}-\frac{1}{\tau}+\frac{\ln\tau}{2\tau^{2}}|=
|\frac{\tau-g(\tau)}{\tau g(\tau)}+\frac{\ln\tau}{2\tau^{2}}|=
\]
\[
=|-\frac{\ln\tau}{2\tau g(\tau)}+\frac{\tau+\ln\tau/2-g(\tau)}{\tau g(\tau)}
+\frac{\ln\tau}{2\tau^{2}}|\leq \frac{C}{\tau^{2}}
\]
Inserting this estimate in (\ref{eq:has2prel}), we get
\begin{equation}\label{eq:hestf}
|h(\tau)-\frac{1}{4\tau}+\frac{1}{8}\frac{\ln\tau}{\tau^2}-
\frac{c_{h,1}}{\tau^2}-\frac{1}{16}\frac{\ln^2 \tau}{\tau^{3}}| 
\leq C\frac{\ln\tau}{\tau^{3}}
\end{equation}

\begin{cor}
Consider a Bianchi VIII solution to
(\ref{eq:whsu})-(\ref{eq:constraint}). There is a constant $C$ and
a $T$ such that 
\begin{equation}\label{eq:vestf}
|v(\tau)-\frac{1}{24\tau^{2}}+\frac{\ln\tau}{24\tau^{3}}|
\leq\frac{C}{\tau^{3}}
\end{equation}
for all $\tau\geq T$.
\end{cor}

\textit{Proof}.
By the proof of Proposition \ref{prop:vas1}, we have
\[
|h^{2}(\tau)-\frac{3}{2}v|\leq\frac{C}{\tau^{3}}.
\]
Combining this with (\ref{eq:has2}) yields the conclusion of the
corollary. $\hfill\Box$

It seems reasonable to think that one should be able to obtain more
terms in the expansions, but we will be satisfied at this point.

\begin{cor}
Consider a Bianchi VIII solution to
(\ref{eq:whsu})-(\ref{eq:constraint}). There are constants $c_{u}$ and
$C$ and a $T$ such that 
\begin{equation}\label{eq:uestf}
|u(\tau)+\frac{1}{4\tau}-\frac{\ln\tau}{8\tau^2}-
\frac{c_{u}}{\tau^{2}}+\frac{\ln^{2}\tau}{16\tau^{3}}|
\leq\frac{C\ln\tau}{\tau^{3}},
\end{equation}
for all $\tau\geq T$.
\end{cor}

\textit{Proof}. The result follows by combining (\ref{eq:hestf}),
(\ref{eq:vestf}) and (\ref{eq:conrel}). $\hfill\Box$

In what follows we will not try to obtain as detailed expansions as 
possible, since it is only a matter of work to do so. The interested 
reader is encouraged to calculate the expansions to higher orders.

\begin{prop}\label{prop:ab8}
Consider a non-NUT Bianchi VIII solution to
(\ref{eq:whsu})-(\ref{eq:constraint}). Then there are constants
$\a_{i}$, $i=1,2,3$, $C$ and a $T$ such that
\begin{equation}\label{eq:aotau}
a_{1}(\tau)=
\a_{1}\tau^{1/2}[1+O(\frac{\ln\tau}{\tau})]
\end{equation}
\begin{equation}\label{eq:atttau}
a_{i}(\tau)=\frac{\a_{i}}{\tau^{1/4}}\exp(3\tau/2)
[1+O(\frac{\ln\tau}{\tau})]
\end{equation}
for $i=2,3$ and all $\tau\geq T$.
\end{prop}

\textit{Proof}.
Using (\ref{eq:uestf}), we conclude the existence of a constant
$c_{s+}$ such that 
\begin{equation}\label{eq:spint}
\int_{0}^{\tau}[\frac{1}{2}-\Spp(s)]ds
=\frac{1}{4}\ln\tau+c_{s+}+O(\frac{\ln\tau}{\tau}).
\end{equation}
Thus, if $\a_{1}=\exp[2c_{s+}]$,
\[
a_{1}(\tau)=\exp[\int_{0}^{\tau}2[\frac{1}{2}-\Spp(s)]ds]=
\a_{1}\tau^{1/2}[1+O(\frac{\ln\tau}{\tau})].
\]
Consider the integral
\begin{equation}\label{eq:intexp}
\int_{0}^{\tau}\Sm(s)ds.
\end{equation}
Observe that $\Sm$ is not $L^{1}([0,\infty))$, but the above mentioned
expression will turn out to converge quite rapidly all the same. By
(\ref{eq:whsu}) we conclude that
\[
\frac{\Nt(\tau)}{\Nth(\tau)}=\exp{\psi(\tau)},
\]
where
\[
\psi(\tau)=\ln\frac{\Nt(0)}{\Nth(0)}+\int_{0}^{\tau}4\sqrt{3}\Sm(s)ds.
\]
Since
\[
\left(\frac{\Nt(\tau)}{\Nth(\tau)}-1\right)^{2}\leq\frac{4}{3}
\frac{h(\tau)}{\Nth^{2}(\tau)},
\]
we conclude that
\[
|\psi(\tau)|\leq\frac{C\cdot h^{1/2}(\tau)}{\Nth(\tau)}.
\]
Thus the integral expression (\ref{eq:intexp}) converges 
exponentially. We have
\[
a_{2}(\tau)=\exp[\int_{0}^{\tau}[1+\Spp(s)+\sqrt{3}\Sm(s)]ds]
\]
\[
=\exp(3\tau/2)\exp[\int_{0}^{\tau}
[\Spp(s)-\frac{1}{2}+\sqrt{3}\Sm(s)]ds].
\]
Thus
\[
a_{2}(\tau)=\frac{\a_{2}}{\tau^{1/4}}\exp(3\tau/2)
[1+O(\frac{\ln\tau}{\tau})].
\]
The argument concerning $a_{3}$ is the same. $\hfill\Box$

\textit{Proof of Theorem \ref{thm:b8a}}.
In order to relate Wainwright Hsu time to proper time, we need
to consider
\[
\int_{0}^{\tau}[1+q(s)]ds=\frac{3}{2}\tau+2\int_{0}^{\tau}\Sm^2(s)ds+
2\int_{0}^{\tau}[\Spp(s)-\frac{1}{2}][\Spp(s)+\frac{1}{2}]ds=
\]
\[
=\frac{3}{2}\tau+2\int_{0}^{\tau}\Sm^2(s)ds+
2\int_{0}^{\tau}[\Spp(s)-\frac{1}{2}]ds+
2\int_{0}^{\tau}[\Spp(s)-\frac{1}{2}]^{2}ds.
\]
Let us start by considering the integral involving $\Sm$. Let
$[\tau_{1},\tau_{2}]$ be an interval such that (\ref{eq:twopi})
is fulfilled. Since we have the approximation (\ref{eq:xvar}),
we get
\[
\int_{\tau_{1}}^{\tau_{2}}\Sm^{2}(s)ds=
h(\tau_{1})\int_{\tau_{1}}^{\tau_{2}}\cos^{2}\xi(\tau)d\tau+
h(\tau_{1})\epsilon_{1}=
\frac{1}{2}h(\tau_{1})(\tau_{2}-\tau_{1})+h(\tau_{1})\epsilon_{2}=
\]
\[
=\frac{1}{2}\int_{\tau_{1}}^{\tau_{2}}h(s)ds+h(\tau_{1})\epsilon_{3}
\]
where $|\epsilon_{i}|\leq C/(\Nt+\Nth)^{2}$,
$i=1,2,3$. In order to obtain this expression, we have also used
(\ref{eq:sinkv}) and (\ref{eq:rvar}). Consequently, there is a 
constant $c_{s-}$ such that 
\[
\phi(\tau)=\int_{0}^{\tau}\Sm^{2}(s)ds-\frac{1}{2}\int_{0}^{\tau}h(s)ds
\rightarrow c_{s-}.
\]
Furthermore,
\[
|\phi(\tau)-c_{s-}|\leq Ce^{-\tau}.
\]
Combining this estimate with (\ref{eq:hestf}), we get the 
conclusion
\[
2\int_{0}^{\tau}\Sm^{2}(s)ds=\frac{1}{4}\ln\tau+c_{s-,1}+
O(\frac{\ln\tau}{\tau}).
\]
Using (\ref{eq:uestf}), we also have
\[
2\int_{0}^{\tau}[\Spp(s)-\frac{1}{2}]ds+
2\int_{0}^{\tau}[\Spp(s)-\frac{1}{2}]^{2}ds=
\]
\[
=-\frac{1}{2}\ln\tau+c_{1}+O(\frac{\ln\tau}{\tau}).
\]
In consequence,
\begin{equation}\label{eq:qint}
\int_{0}^{\tau}[1+q(s)]ds=
\frac{3}{2}\tau-\frac{1}{4}\ln\tau+c_{q,1}+
O(\frac{\ln\tau}{\tau}).
\end{equation}
By (\ref{eq:tp}), we conclude that 
\[
\frac{1}{\theta(\tau)}=\frac{\a_{\theta}}{\tau^{1/4}}e^{3\tau/2}
[1+O(\frac{\ln\tau}{\tau})].
\]
Using (\ref{eq:ttau}), we conclude that 
\begin{equation}\label{eq:tb8}
t(\tau)=\frac{2\a_{\theta}}{\tau^{1/4}}e^{3\tau/2}
[1+O(\frac{\ln\tau}{\tau})].
\end{equation}
This implies that $\tau=2\ln t/3+O(\ln\ln t)$ so that 
\[
\tau^{1/2}=(\frac{2}{3}\ln t)^{1/2}[1+O(\frac{\ln\ln t}{\ln
t})]^{1/2}=(\frac{2}{3}\ln t)^{1/2}[1+O(\frac{\ln\ln t}{\ln
t})].
\]
Combining this with Proposition \ref{prop:ab8}, we conclude that
\[
a_{1}(t)=\a_{1}(\frac{2}{3}\ln t)^{1/2}[1+
O(\frac{\ln\ln t}{\ln t})]
\]
and
\[
a_{i}(t)=\frac{\a_{i}}{2\a_{\theta}}t[1+
O(\frac{\ln\ln t}{\ln t})]
\]
for $i=2,3$. By renaming the $\a_{i}$ we get the conclusion of the
theorem except for the last statement. Note that 
\[
\Nt=(\Nt\Nth)^{1/2}+\Nt^{1/2}\frac{\Nt-\Nth}{\Nt^{1/2}+\Nth^{1/2}}.
\]
The second term tends to zero and will for this reason not be of 
interest to us. Combining (\ref{eq:spint}) and (\ref{eq:qint})
we get 
\[
\frac{(\Nt\Nth)^{1/2}(\tau)}{(\Nt\Nth)^{1/2}(0)}=\exp[\int_{0}^{\tau}
(q+2\Spp)ds]=e^{c}\tau^{-3/4}e^{3\tau/2}[1+O(\frac{\ln\tau}{\tau})].
\]
Consequently, there is a positive constant $c_{0}$ such that 
\begin{equation}\label{eq:ntas}
\Nt(\tau)=c_{0}\tau^{-3/4}e^{3\tau/2}[1+O(\frac{\ln\tau}{\tau})].
\end{equation}
Since 
\[
\frac{a_{2}}{a_{3}}=\left(\frac{\Nth(0)}{\Nt(0)}\right)^{1/2}
\left(\frac{\Nt}{\Nth}\right)^{1/2}
\]
by (\ref{eq:aitau}) and (\ref{eq:whsu}), it is of interest to note 
that 
\[
|\left(\frac{\Nt}{\Nth}\right)^{1/2}-1|\leq
|\frac{\Nt}{\Nth}-1|\leq\frac{2}{\sqrt{3}}\frac{h^{1/2}}{\Nth}
\leq C\frac{\tau^{-1/2}}{\tau^{-3/4}e^{3\tau/2}}\leq\frac{C}{t},
\]
where we have used (\ref{eq:ntas}), (\ref{eq:hestf}) and 
(\ref{eq:tb8}). The theorem follows. $\hfill\Box$

\section{The isometry group of Bianchi VIII initial data}
\label{section:b8i}

In the disc model of Hyperbolic space, the underlying manifold is 
the open unit disc $D$, and the metric is given by 
\[
g_{D}=\frac{4}{(1-x^{2}-y^{2})^{2}}g_{0}
\]
where $g_{0}$ is the standard Euclidean metric on $\mathbb{R}^{2}$.

\begin{lemma}\label{lemma:iso}
Consider an orientation preserving isometry $\phi$ of $(D,g_{D})$, i.e. an
orientation preserving diffeomorphism of $D$ such that
\[
\phi^{*}g_{D}=g_{D}.
\]
Then there is
an $\alpha\in\mathbb{R}$ and a $z_{0}\in D$ such that 
\begin{equation}\label{eq:diso}
\phi(z)=\frac{e^{i\alpha}z+z_{0}}{1+\bar{z}_{0}e^{i\alpha}z}.
\end{equation}
\end{lemma}

\textit{Proof}. The form of the metric implies that $\phi$
preserves angles in the Euclidean sense of the word, at least up
to a sign. The condition that $\phi$ be orientation preserving ensures that 
angles are preserved. As a consequence, $\phi$ must be a biholomorphic
map from $D$ to itself, i.e. a holomorphic map with a holomorphic
inverse. Let
\[
\psi(z)=\frac{z-z_{0}}{1-\bar{z}_{0}z}
\]
where $z_{0}=\phi(0)$. Then 
$f=\psi\circ\phi$ is biholomorphic and $f(0)=0$. By the chain rule
\[
f'(0)(f^{-1})'(0)=1
\]
and by the Schwarz lemma,
\[
|f'(0)|\leq 1,\ \ \mathrm{and}\ \  |(f^{-1})'(0)|\leq 1.
\]
In consequence $|f'(0)|=1$, and the Schwarz lemma yields the
conclusion 
\[
f(z)=e^{i\alpha}z
\]
for some real number $\alpha$. The conclusion follows. $\hfill\Box$

Let $H^{2}=\{(x,y)\in\mathbb{R}^{2}|\ y>0\}$ be the upper half plane.
The map 
\begin{equation}\label{eq:isohd}
\phi_{HD}(z)=\frac{z-i}{z+i}
\end{equation}
defines a biholomorphic map from $H^{2}$ to $D$. If we give $H^{2}$ the
Riemannian metric
\[
g_{H}=\frac{1}{y^{2}}g_{0},
\]
then $\phi_{HD}$ is an isometry. We can of course identify the group
of isometries of $(H^{2},g_{H})$ with the group of isometries of
$(D,g_{D})$ using $\phi_{HD}$. Using this observation together with
Lemma \ref{lemma:iso}, we conclude that the orientation preserving
isometries of $(H^{2},g_{H})$ are the maps of the form
\begin{equation}\label{eq:uhp}
\phi(z)=\frac{az+b}{cz+d}
\end{equation}
where $a,b,c,d\in\mathbb{R}$ and $ad-cb=1$.  Obviously, we have a map 
from $\mathrm{Sl}(2,\mathbb{R})$ onto the orientation preserving isometry
group. This map is a homomorphism, and we have a smooth Lie group
action of $\mathrm{Sl}(2,\mathbb{R})$ on the left of $H^{2}$. Since the 
kernel is generated by $-\mathrm{Id}$, we get a smooth action of 
$\mathrm{PSl}(2,\mathbb{R})$ on $H^{2}$. The action of this group is
effective. We can thus identify the group of orientation preserving
isometries of hyperbolic space with $\mathrm{PSl}(2,\mathbb{R})$. 
 The action of
$\mathrm{PSl}(2,\mathbb{R})$ on $H^{2}$ can be extended to an action on
$UH^{2}$, the unit tangent bundle of hyperbolic space, by for each group
element $g$ considering the push forward of the corresponding map 
(below, the push forward of the action of a Lie group element $g$ on
hyperbolic space will be denoted by $g_{*}$). In
this way we get a smooth Lie group action on $UH^{2}$ on the left which is 
free and transitive, and if we fix an element $e\in UH^{2}$, the map
obtained by letting the Lie group act on this element has non-degenerate
derivative at each Lie group element. Consequently, after having
chosen an identity element $e\in UH^{2}$, we have a diffeomorphism from 
$\mathrm{PSl}(2,\mathbb{R})$ to $UH^{2}$ given by 
$\phi(g)=g_{*}e$. It will be convenient to use different representations
of hyperbolic space at different times. Conclusions that have been
obtained using one representation can trivially be carried over to the 
others and we will do so without comment. For instance, we will
consider the map defined by (\ref{eq:diso}) to be an element of $\psl$
and consider $UH^{2}$ and $UD$ to be the same. One can define a map 
$\rho: \mathbb{R}\times D\rightarrow UD$ by 
\[
\rho(\alpha,z)=\cos\alpha e_{1,z}+\sin\alpha e_{2,z}
\]
where $e_{i}$ is the unit vector in the direction of $\partial_{i}$.
This defines a covering map of $UD$, and by going to the quotient,
one obtains a diffeomorphism from $S^{1}\times D$ to $UD$. Locally,
$\rho$ defines what we shall refer to as $\alpha xy$-coordinates.
Observe that the above means that $\tsl$ is $\mathbb{R}^{3}$ topologically.

We have a diffeomorphism from $\psl$ to $UD$. It can be used to give 
$UD$ a group structure such that the diffeomorphism becomes
a Lie group isomorphism. Let $e=e_{1,0}\in UD$ be the identity element.
If $\phi_{i}$ are orientation preserving isometries of $(D,g_{D})$ for
$i=1,2$, we have
\begin{equation}\label{eq:prod}
(\phi_{1*}e)(\phi_{2*}e)=(\phi_{1}\phi_{2})_{*}e.
\end{equation}
Since $\rho$ is a covering map, there is a unique way of making 
$\mathbb{R}\times D$ a Lie group such that $\rho$ becomes a homomorphism,
assuming one has chosen an identity element, see e.g. \cite{lie}. 
Let $(0,0)\in \mathbb{R}\times D$ be the identity. Then the fact that 
$\rho$ is a covering map and a homomorphism can be used to conclude that 
\begin{equation}\label{eq:twopro}
(2\pi t,0)\cdot (\alpha,z)=(\alpha+2\pi t,e^{i2\pi t}z),\ \ \
(\alpha,z)\cdot (2\pi t,0)=(\alpha+2\pi t,z).
\end{equation}
In particular, $(2\pi n,0)$ commutes with all elements of $\tsl$ 
and corresponds to a translation by $2\pi n$. These translations 
constitute the kernel of $\rho$. 

We are interested in finding compactifications of Bianchi VIII initial
data, and in that context, the following observation is relevant. Note
that this contradicts a statement in \cite{tan}. However, as was noted in
\cite{tan2}, the statement found in \cite{tan} is incorrect. 

\begin{lemma}\label{lemma:ce}
For every $p\in \mathbb{N}$ with $p>1$, there is a subgroup
$\Xi_{p}$ of $\tsl$ such that $\Xi_{p}$, considered as a group of
diffeomorphisms acting on the left, is a free and properly
discontinuous group of diffeomorphisms such that 
\[
\tsl/\Xi_{p}\cong U\Sigma_{p},
\]
where $\Sigma_{p}$ is the compact orientable $2$-manifold of genus $p$,
and $U\Sigma_{p}$ is the unit tangent bundle of this surface 
with respect to a suitable hyperbolic metric. Here $\cong$
symbolizes the existence of a diffeomorphism.
\end{lemma}
\textit{Remark}. The above statement is of course trivial. Something
which is far from trivial 
is to classify all the subgroups $\Gamma$ of $\tsl$ that are free and
properly discontinuous when acting on the left and yield compact quotients.
This is done in \cite{ray}, where the corresponding topological spaces
are also described. 

\textit{Proof}. Let $\phi:\psl\rightarrow UH$ be defined
by $\phi(g)=g_{*}e$, where $e$ is a fixed element of $UH$. Then
$hg=\phi^{-1}[h_{*}\phi(g)]$,
so that a subgroup $\Gamma$ of $\psl$ acting on the left is a 
free and properly discontinuous group of diffeomorphisms of $\psl$ 
if and only if the group $\Gamma_{*}=\{ g_{*}|g\in \Gamma\}$ is a free
and properly discontinuous group of diffeomorphisms of $UH^{2}$. In that
case, $\phi$ defined above yields a diffeomorphism
\[
\hat{\phi}:\mathrm{PSl}(2,\mathbb{R})/\Gamma\rightarrow UH^{2}/\Gamma_{*}.
\]
For every $p>1$, there is a  subgroup $\Gamma_{p}$ of 
$\mathrm{PSl}(2,\mathbb{R})$, such that $\Gamma_{p}$ acting on $H^{2}$
is a free and properly discontinuous group of orientation preserving
isometries of $H^{2}$ such that 
$H^{2}/\Gamma_{p}$ is diffeomorphic to $\Sigma_{p}$, the compact orientable
$2$-manifold with genus $p$. The push forward of the
projection map $\pi_{p} :H^{2}\rightarrow \Sigma_{p}$ defines a smooth map 
$\pi_{p*}: UH^{2}\rightarrow U\Sigma_{p}$, where $U\Sigma_{p}$ is the 
unit tangent bundle of $\Sigma_{p}$ with respect to the natural hyperbolic 
metric induced by taking the quotient. The map
$\pi_{p*}$ identifies $v_{1}$ and $v_{2}$ if and only if
$v_{2}=h_{*}v_{1}$ for $h_{*}\in \Gamma_{p*}=\{ h_{*}|
h\in \Gamma_{p}\}$. Since $\Gamma_{p*}$ is a free and properly 
discontinuous group of diffeomorphisms on $UH^{2}$, the map $\pi_{p*}$ 
defines an isometry
\[
\psi :UH^{2}/\Gamma_{p*}\rightarrow U\Sigma_{p}.
\] 
By the above correspondence, we get a covering projection
\[
\pi_{\Sigma_{p}}:\mathrm{PSl}(2,\mathbb{R})\rightarrow U\Sigma_{p}
\]
where the covering transformations are the left translations by
elements of $\Gamma_{p}$. Furthermore, we have the covering projection
\[
\pi_{\mathrm{PSl}}:\tsl
\rightarrow\mathrm{PSl}(2,\mathbb{R}).
\]
Observe that $\Xi_{p}=\pi_{\mathrm{PSl}}^{-1}\Gamma_{p}$ is a subgroup of
$\tsl$ since the projection is a homomorphism. Composing two
covering projections, one does not necessarily get a covering
projection, but in our situation, this is the case, see for instance
Spanier \cite{spanier}. One can also check that the group of covering
transformations consists of the subgroup $\Xi_{p}$ acting on the 
left on $\tsl$. The lemma follows. $\hfill\Box$

The discussion below, until and including the proof of Lemma 
\ref{lemma:asg}, is essentially taken from \cite{scott}.
It will be necessary in order to prove Theorem \ref{thm:seifert}.
Note that one form of isometry is a reflection in a geodesic. By
this we mean a map which takes the geodesic to itself and maps every
vector $v$ perpendicular to the geodesic to $-v$. Explicitly, we have the
reflections
\[
\phi(x,y)=(2x_{0}-x,y),\ \ \
\phi(z)=\frac{r^{2}}{\bar{z}-a}+a,
\]
where $r,a,x_{0}\in\mathbb{R}$ and $r>0$, in the upper half plane. 
Given the representation (\ref{eq:uhp}) of the orientation preserving
isometries, one can check that any orientation preserving isometry can 
be represented as a product of two reflections in geodesics. In fact,
two reflections in straight geodesics gives the translations, two reflections
in circular geodesics yield the maps of the form $az+b$ for $a,b\in\mathbb{R}$
and $a>0$, $a\neq 1$. Finally, a combination of a reflection in a straight
geodesic and a reflection in a circular geodesic yields the remaining
isometries. Note that if $\alpha$
is a reflection in a geodesic $\gamma$ and $\phi$ is an isometry,
then $\phi \alpha \phi^{-1}$ is a reflection in the geodesic
$\phi \gamma$. Furthermore, there are three distinct combinations
of geodesics. i) The geodesics intersect in an interior point of 
hyperbolic space. ii) The geodesics intersect on the boundary when
viewed in the disc model. iii) The geodesics do not intersect
in the interior or on the boundary. The respective names for the 
corresponding isometries are \textit{rotation}, \textit{parabolic} 
isometry and \textit{hyperbolic} isometry. Let us describe the 
possibilities in somewhat greater detail. Consider a rotation $\psi$. 
Let $\phi$ be an isometry which takes the intersection point of 
the two geodesics to the origin in the disc model. The geodesics then
become straight lines through the origin and $\phi\psi\phi^{-1}$ is
a rotation in the Euclidean sense of the word. The angle of rotation
is twice the angle between the two geodesics. Thus non-trivial
rotations leave one
interior point of hyperbolic space fixed but no other points. Two 
geodesics which intersect at infinity become, after applying an isometry,
two straight lines in the upper half plane. In the upper half plane,
the resulting isometry is translation in the $x$-direction twice the
Euclidean distance between the lines. It leaves
exactly one point on the boundary of hyperbolic space fixed, but no
point in the interior. Let $\psi$ be a hyperbolic isometry. The two 
geodesics can be assumed to be a straight line
and a circle with center on the real line in the upper half plane. One
sees that there must be a unique geodesic $\gamma$ intersecting
each of the two geodesics at straight angles. Observe that both the
reflections map $\gamma(\mathbb{R})$ to itself, so that the 
composition maps the geodesic to itself and preserves the orientation.
Let $\phi$ be
an isometry taking $\gamma$ to the real line intersected with the 
disc. Then $\phi \psi \phi^{-1}$ is an isometry of the disc
model taking the real line intersected with the disc into itself. 
Since every isometry of the disc can be written in the form 
\[
\chi(z)=\frac{e^{i\alpha}z+z_{0}}{1+\bar{z}_{0}e^{i\alpha}z},
\]
we get the conclusion that $z_{0}$ has to be real and $e^{i\alpha}=\pm
1$. Since $\chi$ also preserves the orientation of the real line,
$e^{i\a}=1$. Thus
\begin{equation}\label{eq:hyp}
\chi(z)=\frac{z+t}{1+tz},
\end{equation}
where $t\in\mathbb{R}$. For $t\neq 0$, such a map fixes two points on
the boundary but no other points. Consider an orientation
reversing isometry $\psi$ of the unit disc model. Observe that the
isometry can be extended to a holomorphic map from a neighbourhood of the
closed disc into the complex plane. Furthermore, the map will have a
non-degenerate derivative and will still map orthogonal vectors,
with respect to the Euclidean metric, to orthogonal vectors. Since
$\psi$ maps the boundary to itself, a vector normal to the boundary
will be mapped to a non-zero multiple of itself. Since the map also
maps the interior of the disc to itself, it has to be a positive
multiple. Since $\psi$ is orientation reversing, the map from the 
boundary to itself thus has to be orientation reversing. Thus $\psi$
has to have two fixed points on the boundary. 

It will be of 
interest to find, given a non-trivial orientation preserving isometry
$\phi$, the group $C(\phi)$ of isometries commuting with $\phi$.

\begin{lemma}\label{lemma:asg}
Let $\phi$ be a non-trivial orientation preserving isometry of
hyperbolic space. 
\begin{enumerate}
\item If $\phi$ is a rotation, then there is an orientation preserving
isometry $\chi$ such that $\chi C(\phi)\chi^{-1}$ is the group of
rotations around the origin in the disc model.
\item If $\phi$ is a hyperbolic isometry, then there is an orientation 
preserving isometry
$\chi$ such that $\chi C(\phi)\chi^{-1}$ is generated by maps of the 
form (\ref{eq:hyp}) and a reflection in the real axis. 
\item If $\phi$ is parabolic, then there is an orientation preserving
isometry $\chi$ such
that $\chi C(\phi)\chi^{-1}$ is the group of translations by  real
numbers in the upper half plane.
\end{enumerate}
\end{lemma}

\textit{Proof}. Note that if $\phi$ and $\psi$ are two commuting maps, then 
$\psi=\phi\psi\phi^{-1}$ fixes $\phi[\mathrm{fix}(\psi)]$, 
where $\mathrm{fix}(\psi)$ is the set of fixed points of $\psi$.
Thus $\phi$ leaves $\mathrm{fix}(\psi)$ invariant. 
If $\phi$ is a rotation, then it fixes a point $z$ of the interior,
but no other point. Thus any $\psi\in C(\phi)$ must be a rotation
around $z$. If $\chi$ takes $z$ to the origin of the disc
model, we get the conclusion that $\chi C(\phi)\chi^{-1}$ 
coincides with the Euclidean rotations around the origin. 
If $\phi$ is hyperbolic, let $\chi$ be such that $\chi\phi\chi^{-1}$
is of the form (\ref{eq:hyp}). Since 
$\chi C(\phi)\chi^{-1}=C(\chi\phi\chi^{-1})$, we can assume $\phi$ to
be of the form (\ref{eq:hyp}). Note that $C(\phi)$ contains all
isometries of this form plus the reflection in the real axis. We
wish to prove that this is all there is. If $\psi\in C(\phi)$ and
$\psi$ is orientation preserving, then $\psi$ fixes the points $\pm 1$
or interchanges them. If it fixes $\pm 1$, it must be of the form
(\ref{eq:hyp}). If it interchanges $\pm 1$, we can compose $\psi$ 
with a rotation to get an orientation preserving isometry fixing
$\pm 1$. Thus the composition must be of the form (\ref{eq:hyp})
whence the non-trivial rotation must belong to $C(\phi)$ due to the
group properties of this set. This is impossible. Given an orientation
reversing isometry $\psi\in C(\phi)$, we compose it with a reflection
in the real line. By the above, the composition is of the form
(\ref{eq:hyp}). Finally, consider a non-trivial parabolic isometry
$\phi$. We can assume that $\phi$ is a non-trivial translation by a
real number in the upper half plane. Since $\phi$ only leaves one
point on the boundary fixed it cannot commute with an orientation
reversing isometry. By the representation (\ref{eq:uhp}) one can
check that the orientation preserving isometries that commute with a
non-trivial translation by a real number are translations by a real
number. $\hfill\Box$

Let us construct a basis for the Lie algebra of $\psl$. By the above,
we can identify $\psl$ with $UD$. Let the unit vectorfields $e_{1}$ and
$e_{2}$ be the normalizations of $\partial_{x}$ and $\partial_{y}$,
and let us identify the identity element of the group with the vector
$e_{1}$ at zero. Consider the three curves in $UD$ defined by 
$\gamma_{1}(t)=e_{1,0}\cos (t)+e_{2,0}\sin (t)$
$\gamma_{2}(t)=e_{1,t/2}$ and
$\gamma_{3}(t)=e_{1,\mathrm{i}t/2}$. The derivatives of
these curves for $t=0$ define three tangent vectors at the identity
which we will call $E_{i,0}$.
In terms of $\alpha xy$-coordinates close to $(0,0,0)$, we have 
\[
\gamma_{1}(t)=(t,0,0),\
\gamma_{2}(t)=(0,t/2,0)\ \mathrm{and}\
\gamma_{3}(t)=(0,0,t/2),
\]
so that the tangent vectors to the different curves form a basis at
the identity. We denote the corresponding left invariant vector
fields $E_{i,p}$ for $p\in UD$. After some computations, one finds
\begin{equation}\label{eq:canb}
E_{1}=\partial_{\alpha},\ \ 
E_{2}=\cos\alpha e_{1}+\sin\alpha e_{2}+f\partial_{\alpha},\ 
E_{3}=-\sin\alpha e_{1}+\cos\alpha e_{2}+g\partial_{\alpha},
\end{equation}
where
\[
f=y \cos \alpha -x \sin\alpha ,\ \ \
g=-(x \cos\alpha +y \sin\alpha ).
\]
Note that the $E_{i}$ also constitute a left invariant basis on
$\mathbb{R}\times D\cong \tsl$. The structure constants 
$\gamma_{ij}^{k}$, determined by
\[
[E_{i},E_{j}]=\gamma_{ij}^{k}E_{k},
\]
can be computed to be $\gamma_{ij}^{k}=\epsilon_{ijl}n^{lk}$
where $n=\mathrm{diag}(-1,1,1)$, which is the defining characteristic of 
Bianchi VIII. Later, it will be important to know that if we change the 
basis according to 
\[
\hat{E}_{i}=(A^{-1})_{i}^{\ k}E_{k},
\]
we get, for the structure constants
$\hat{\gamma}_{ij}^{k}=\epsilon_{ijl}\hat{n}^{lk}$ corresponding
to $\hat{E}_{i}$,
\begin{equation}\label{eq:ntrans}
\hat{n}=\frac{1}{\det A}{}^{t}AnA.
\end{equation}

Before we go on, let us note the following. If $G_{i}$ are two
simply connected Lie groups and $\phi$ is a Lie algebra isomorphism
between their Lie algebras, there is an analytic group isomorphism
\[
\psi:G_{1} \rightarrow G_{2}
\]
such that
\[
\exp\circ\phi=\psi\circ\exp,
\]
see e.g. \cite{lie}.

Let $g$ be a left invariant metric and $k$ a left invariant 
symmetric and covariant $2$-tensor. We assume that they are diagonal
with respect to the basis $E_{i}$, but $g$ need not be
orthonormal with respect to this basis. If we rescale the 
$E_{i}$ by positive numbers, we get an orthonormal basis 
$\tilde{E}_{i}$. With respect to this basis, $n$ is still diagonal 
with $n_{1}<0$ and 
$n_{2},n_{3}<0$. We will denote the diagonal elements by
$k_{1}=k(\tilde{E}_{1},\tilde{E}_{1})$ etc. We are interested in the
subgroup $\mathcal{D}(g,k)$ of the group of diffeomorphisms 
$\phi:\tsl\rightarrow \tsl$ such that 
\begin{equation}\label{eq:piso}
\phi^{*}k=k\ \ \ \mathrm{and} \ \ \ \phi^{*}g=g. 
\end{equation}
First of all, left translations are in $\mathcal{D}(g,k)$. However,
$\mathcal{D}(g,k)$ always contains three more elements which we now
define. Consider a matrix $A$ which is diagonal with two minus signs
and one plus sign. Due to (\ref{eq:ntrans}), $A$ defines a Lie Algebra
isomorphism. As noted above, this yields a Lie group isomorphism
which also satisfies (\ref{eq:piso}). Explicitly, these isometries
are defined by 
\begin{equation}\label{eq:tthree}
\phi_{1}(\alpha,z)=(\alpha,-z), \ \
\phi_{2}(\alpha,x,y)=(-\alpha,-x,y),\ \
\phi_{3}(\alpha,x,y)=(-\alpha,x,-y).
\end{equation}
We will denote the group of diffeomorphisms
generated by left translations and these additional isometries by
$\mathcal{D}_{s}$. If $k_{2}=k_{3}$ and $n_{2}=n_{3}$, the matrix
$A$ leaving $\tilde{E}_{1}$ invariant and rotating 
$\tilde{E}_{2},\tilde{E}_{3}$ by some angle
$\theta$ is also a Lie Algebra isomorphism. It thus defines
a Lie Group isomorphism satisfying (\ref{eq:piso}). 
Adding these diffeomorphisms to $\mathcal{D}_{s}$, we get a group of 
diffeomorphisms we will refer to as $\mathcal{D}_{e}$ ($s$ for
standard and $e$ for exceptional).

Let us make a few observations concerning $\mathcal{D}_{s}$ and 
$\mathcal{D}_{e}$. 
\begin{lemma}\label{lemma:isom}
For any $\theta\in\mathbb{R}$, the map $T_{\theta}$ defined by
$T_{\theta}(\alpha,z)=(\alpha+\theta,z)$ is in $\mathcal{D}_{e}$
and the map defined by $\theta\mapsto T_{\theta}$ defines an
injective homomorphism from $\mathbb{R}$ to $\mathcal{D}_{e}$.
There is a map from $\mathcal{D}_{e}$ to maps from $D$ to itself
defined by
\begin{equation}\label{eq:pdef}
p[\phi](z)=\pi_{2}[\phi(0,z)],
\end{equation}
where $\pi_{2}$ is the projection to the second factor. Then $p$ is
a surjective homomorphism from $\mathcal{D}_{e}$ to 
$\mathrm{Isom}(H^{2})$, and 
\begin{equation}\label{eq:indep}
p[\phi](z)=\pi_{2}[\phi(\alpha,z)].
\end{equation}
Furthermore, we have the exact sequence 
\begin{equation}\label{eq:exact}
0\longrightarrow \mathbb{R}\longrightarrow \mathcal{D}_{e}
\stackrel{p}{\longrightarrow}\mathrm{Isom}(H^{2})\longrightarrow \{ e\}.
\end{equation}
Translations by $n\pi$ are in $\mathcal{D}_{s}$ and the map
defined by $n\mapsto T_{n\pi}$ defines an injective homomorphism
from $\mathbb{Z}$ to $\mathcal{D}_{s}$ and the following sequence
is exact
\[
0\longrightarrow \mathbb{Z}\longrightarrow \mathcal{D}_{s}
\stackrel{p}{\longrightarrow}\mathrm{Isom}(H^{2})\longrightarrow \{ e\}.
\]
Finally $\mathcal{D}_{e}$ can be given the structure of a Lie group
with exactly two connected components
such that the connected component of the identity $\mathcal{D}_{e,o}$ 
is isomorphic
with $\mathbb{R}\times \tsl/K$, where $K$ is the
subgroup generated by $[2\pi ,(-2\pi ,0)]$. If we give
$\mathrm{Isom}(H^{2})$ a Lie group structure by identifying the 
orientation preserving part with $\psl$, then $p$ is a smooth map
taking open sets to open sets. Finally, translations 
commute with elements of $\mathcal{D}_{e,o}$. 
\end{lemma}
\textit{Remark}. Note that if we would have $\tsl$ instead of
$\mathcal{D}_{s}$ we would have the exact sequence
\[
0\longrightarrow \mathbb{Z}\longrightarrow \tsl
\stackrel{p}{\longrightarrow} \psl \longrightarrow \{ e\}.
\]
However, the $\mathbb{Z}$ would correspond to translations by
multiples of $2\pi$ and not of $\pi$. For $t\in\mathbb{R}$ and
$h\in\tsl$, we will sometimes write $t h$ instead of $T_{t}(h)$.

\textit{Proof}.
Consider the diffeomorphism $T_{\theta}$
of $\mathbb{R}\times D$ to itself defined by
$T_{\theta}(\alpha,z)=(\alpha+\theta,z)$. We have
\begin{equation}\label{eq:rot1}
T_{\theta *}E_{1,(\alpha,z)}=E_{1,(\alpha+\theta,z)},\ \
T_{\theta *}E_{2,(\alpha,z)}=\cos \theta E_{2,(\alpha+\theta,z)}-
\sin\theta E_{3,(\alpha+\theta,z)}
\end{equation}
and
\begin{equation}\label{eq:rot2}
T_{\theta *}E_{3,(\alpha,z)}=\sin \theta E_{2,(\alpha+\theta,z)}+
\cos\theta E_{3,(\alpha+\theta,z)}.
\end{equation}
If we put a metric $\tilde{g}$ on $\tsl$ such that $E_{i}$ is an
orthonormal basis, we get the conclusion that $T_{\theta}$ is 
an isometry of $(\tsl,\tilde{g})$. Furthermore, if we compose
$T_{\theta}$ with a left translation corresponding to the element
$(-\theta,0)$ of $\tsl$ we get a map $L_{(-\theta,0)}T_{\theta}$
from $\tsl$ to itself taking the identity to itself and which is 
a rotation by an angle $\theta$ of $(E_{2},E_{3})$ in the counter 
clockwise direction on the tangent space at the identity, cf. 
(\ref{eq:rot1}), (\ref{eq:rot2}) and (\ref{eq:twopro}). As noted
above, there is however a Lie group
isomorphism $\psi_{\theta}$ with exactly this effect on the 
tangent space at the identity. Thus $\psi_{\theta}^{-1}L_{(-\theta,0)}
T_{\theta}$ is an isometry of $\tilde{g}$ taking the identity to the
identity and which is the identity on the tangent space at the
identity. Since $\tsl$ is connected, $T_{\theta}=L_{(\theta,0)}\psi_{\theta}$.
In particular all translations are in $\mathcal{D}_{e}$. 
Translations commute with left translations for the following
reason. By (\ref{eq:twopro}), (\ref{eq:rot1}) and (\ref{eq:rot2}), 
$L_{(\alpha,z)}T_{\theta}$ and $T_{\theta}L_{(\alpha,z)}$ map the 
identity to the same point and have the same effect on the tangent space
at the identity. Let us note the following. The map $\phi_{1}$ in 
(\ref{eq:tthree}) can be written as a combination of a translation by
$\pi$ and a left translation. Thus translations by $\pi$ are in 
$\mathcal{D}_{s}$. Left translations do not commute
with $\phi_{2}$, $\phi_{3}$, but
$\phi_{i}L_{g}=L_{\phi_{i}(g)}\phi_{i}$ for $i=2,3$. Similarly,
for translations, $T_{\theta}\phi_{i}=\phi_{i}T_{-\theta}$.
Finally $\phi_{2}\phi_{3}=\phi_{1}$. Consequently, any element of
$\psi\in\mathcal{D}_{e}$ can be written in the form 
\begin{equation}\label{eq:representation}
\psi=\phi T_{\theta}L_{g},
\end{equation}
where $\phi$ is either $\phi_{2}$ or the identity, $T_{\theta}$ is a
translation and $L_{g}$ is a left translation. 

Let $p$ be defined by (\ref{eq:pdef}). For some $\phi$, we also have
(\ref{eq:indep}). In fact, if $\phi$ is a translation, one of 
(\ref{eq:tthree}) or 
a left translation, this is true. If $\phi=\chi_{1}\chi_{2}$
where $\chi_{1}$ is such a map, we get
\[
p[\phi](z)=\pi_{2}[\chi_{1}\chi_{2}(0,z)]=
p[\chi_{1}](\pi_{2}[\chi_{2}(0,z)])=
p[\chi_{1}]p[\chi_{2}](z).
\]
Applying this observation repeatedly using (\ref{eq:representation}), 
we get the conclusion that 
$p$ is a homomorphism and that (\ref{eq:indep}) holds for all $\phi$. 
Let us compute $p$ if $\phi$ is a left translation. Note that 
$\pi_{2}$ can be factored through $\rho$, the covering map from
$\mathbb{R}\times D$ to $UD$. If $\pi_{D}:UD\rightarrow D$ is the 
projection to the base, we have $\pi_{2}=\pi_{D}\circ \rho$.
We get
\[
p[L_{(\alpha,z)}](\zeta)=\pi_{2}[(\alpha,z)\cdot (0,\zeta)]=
\pi_{D}[\rho(\alpha,z)\cdot \rho(0,\zeta)]=
\pi_{D}[\phi_{*}\rho(0,\zeta)]
=\phi(\zeta),
\]
where $\phi$ is the orientation preserving isometry of $(D,g_{D})$
such that $\phi_{*}e_{1,0}=\rho(\alpha,z)$, cf. (\ref{eq:prod}).
Thus $p[L_{(\alpha,z)}]$ is an isometry of hyperbolic space.
If $\phi$ is one of (\ref{eq:tthree}) it is not so difficult
to compute $p[\phi]$, and if $\phi$ is a translation, one obtains the 
identity. Thus $p$ is a homomorphism from $\mathcal{D}_{e}$ to the 
isometry group of hyperbolic space. Since the image of the left translations
is the orientation preserving isometries and $p[\phi_{2}]$ is orientation
reversing, $p$ is surjective. By the representation (\ref{eq:representation})
we get the conclusion that the kernel of $p$ is the group of translations.
Let us give $\mathrm{Isom}(H^{2})$ the structure of a Lie group by
identifying the orientation preserving part with $\psl$. By our
previous discussions, we can identify $\psl$ with $UD$ by equating
$\phi\in\psl$ with $\phi_{*}e_{1,0}$, the push forward of the unit
vector in the $x$-direction at the origin. With these identifications,
$p$ restricted to $\tsl$ is simply the covering projection $\rho$ from 
$\mathbb{R}\times D$ to $UD$. 

Let $\mathcal{D}_{e,o}$ be the subgroup of $\mathcal{D}_{e}$ generated
by translations and left translations. Note that
$\phi_{2}\mathcal{D}_{e,o}$ is disjoint from $\mathcal{D}_{e,o}$ 
since $p(\mathcal{D}_{e,o})$ consists of orientation preserving
isometries and $p(\phi_{2}\mathcal{D}_{e,o})$ of orientation reversing
isometries. We have a surjective homomorphism from $\mathbb{R}\times
\tsl$ to $\mathcal{D}_{e,o}$, and the kernel $K$ is
generated by $[2\pi,(-2\pi,0)]$. Thus $\mathcal{D}_{e,o}$ can be given
the structure of 
a connected Lie group as stated in the lemma. Using $\phi_{2}$, we can
give $\mathcal{D}_{e}$ the structure of a Lie group with exactly two 
connected components. We can define a map $p':\mathbb{R}\times \tsl
\rightarrow \mathrm{Isom}(H^{2})$ by $p'(t,h)=p[L_{h}]$. By the
observations in the previous paragraph, $p'$ can be identified with
the map from $\mathbb{R}\times\mathbb{R}\times D$ to $UD$ sending
$(t,\a,z)$ to $\rho(\a,z)$. Consequently, $p'$ descends to a smooth map
from the quotient $\mathbb{R}\times\tsl/K$ to $\mathrm{Isom}(H^{2})$
which can be identified as $p$ restricted to the identity component of 
$\mathcal{D}_{e}$. Thus we see that $p$ is a smooth map with respect
to the above mentioned topologies and that it takes open sets to open
sets. $\hfill\Box$

\begin{prop}\label{prop:isom}
Let $(g,k)$ be a left invariant metric and symmetric covariant 
$2$-tensor satisfying Einstein's constraint equations. Assume
further that they are diagonal with respect to the 
basis $E_{i}$. Let $\tilde{E}_{i}$ be an orthonormal basis obtained
by rescaling $E_{i}$ by positive numbers and let the diagonal 
components of $n$ and $k$
with respect to $\tilde{E}_{i}$ be denoted $n_{i}$ and $k_{i}$.
Then if $n_{2}=n_{3}$ and $k_{2}=k_{3}$,
$\mathcal{D}(g,k)=\mathcal{D}_{e}$. Otherwise
$\mathcal{D}(g,k)=\mathcal{D}_{s}$.
\end{prop}

\textit{Proof}. Let $\phi\in\mathcal{D}(g,k)$.
Given a vectorfield $X$, we can define a vectorfield by
\[
(\phi_{*d} X)_{p}=\phi_{*}X_{\phi^{-1}(p)}.
\]
We will use the notation
\[
E_{i}'=\phi_{*d}\tilde{E}_{i}\ \ \
\mathrm{and}\ \ \
A_{ik}=<E_{i}',\tilde{E}_{k}>.
\]
Note that $A$ is an orthogonal matrix, but that it is not yet
clear that it is constant.
The strategy of determining $\mathcal{D}(g,k)$ will be to 
find a $\psi$ in $\mathcal{D}_{e}(g,k)$ (if $n_{2}=n_{3}$ and 
$k_{2}=k_{3}$) or in $\mathcal{D}_{s}(g,k)$ such that 
$\psi[\phi(e)]=e$ and $\psi_{*}\phi_{*}$ is the identity on
when restricted to the tangent space at the identity element. Since
$\psi\circ\phi$ is an isometry and $\tsl$ is connected, this would
imply $\psi\circ\phi=\mathrm{Id}$. Note that it is enough to prove 
that there is a $\psi$ in the relevant group such that
$\psi_{*d}\phi_{*d}\tilde{E}_{i}=\tilde{E}_{i}$ for all $i$, since a left
translation takes $\psi[\phi(e)]$ to the identity element and leaves
the $E_{i}$ unchanged. Observe finally that if $\tilde{E}_{i}$ is 
a basis of NUT type, then the scale factor relating $E_{2}$ and 
$\tilde{E}_{2}$ is the same as the scale factor relating $E_{3}$ and
$\tilde{E}_{3}$ so that a rotation of $E_{2}$ and $E_{3}$ corresponds
to a rotation of $\tilde{E}_{2}$ and $\tilde{E}_{3}$ in this case. 

Since $\phi$ is an isometry, it preserves the Levi-Civita connection
so that 
\[
\nabla_{\phi_{*d}X}\phi_{*d}Y=\phi_{*d}(\nabla_{X}Y).
\]
Note that 
\begin{equation}\label{eq:seg}
\mathrm{Ric}(E_{i}',E_{j}')=\mathrm{Ric}(\tilde{E}_{i},\tilde{E}_{j})\ \
\mathrm{and}\ \ \ k(E_{i}',E_{j}')=k(\tilde{E}_{i},\tilde{E}_{j})
\end{equation}
since $\phi$ satisfies (\ref{eq:piso}). In principle, the
right and the left hand sides should be evaluated at different points, but
since the right hand sides are constant, it is not so important to keep
this in mind. Let us define
\[
\tilde{R}(X)=\sum_{m}\mathrm{Ric}(X,\tilde{E}_{m})\tilde{E}_{m}\ \ \
\mathrm{and}\ \ \
\tilde{k}(X)=\sum_{m}k(X,\tilde{E}_{m})\tilde{E}_{m}
\]
Observe that these operators are independent of the orthonormal basis
used to define them. One can compute that 
\[
\mathrm{Ric}(\tilde{E}_{i},\tilde{E}_{j})=2n_{i}^{\ k}n_{kj}-
\mathrm{tr}(n)n_{ij}-n^{kl}n_{kl}\delta_{ij}+\frac{1}{2}
[\mathrm{tr}(n)]^{2}\delta_{ij}.
\]
Here, indices are raised and lowered with $\delta_{ij}$.
Thus $\tilde{E}_{i}$ and $E_{i}'$ are eigenvalues of $\tilde{R}$ 
with the same eigenvalues, and similarly for $\tilde{k}$, 
due to (\ref{eq:seg}).
Note that since $k$ and $\mathrm{Ric}$ are symmetric, eigenvectors
of $\tilde{R}$ and $\tilde{k}$ with different eigenvalues are
orthogonal. By the above it is clear that $R_{ij}$
is diagonal, and we will refer to the diagonal components as $R_{i}$.
Observe that
\begin{equation}\label{eq:eigen}
R_{i}=2n_{i}^{2}-\mathrm{tr}(n)n_{i}-n^{j}n_{j}+\frac{1}{2}
[\mathrm{tr}(n)]^{2}.
\end{equation}
The assumption $R_{2}=R_{3}$ leads to the conclusion that
\[
n_{2}^{2}-(n_{1}+n_{3})n_{2}=n_{3}^{2}-(n_{1}+n_{2})n_{3}
\]
whence
\[
n_{2}(n_{2}-n_{1})=n_{3}(n_{3}-n_{1}).
\]
Observe that since $n_{1}<0$ and $n_{2}$ and $n_{3}$ are
positive, this implies that $n_{2}=n_{3}$. Thus $R_{2}=R_{3}$
is equivalent to $n_{2}=n_{3}$. Assuming $R_{1}=R_{2}$
leads to the conclusion
\[
n_{1}(n_{1}-n_{3})=n_{2}(n_{2}-n_{3}).
\]
Observe that the left hand side is positive, so that $n_{2}>n_{3}$
is a consequence. In particular, all three eigenvalues of Ricci
cannot be equal. Consider now the following cases of initial data.

i) If all eigenvalues of the Ricci tensor are different, we conclude that
$A$ is diagonal with entries $\pm 1$. 
Since $\phi$ is an isometry, the structure constants corresponding
to $E_{i}'$ must coincide with those corresponding to $\tilde{E}_{i}$. 
Considering (\ref{eq:ntrans}) we get the conclusion that $A$ must
be the identity or have two minus signs and one plus sign on the 
diagonal. In other words, $\phi\in \mathcal{D}_{s}$. 

ii) If $n_{2}=n_{3}$, then $A_{i1}=A_{1i}=0$ for all $i\neq 1$,
and $A_{11}=\pm 1$. If $k_{2}\neq k_{3}$, then $A$ has to be diagonal
with diagonal entries $\pm 1$ and we have the same situation as in
the previous case. If $k_{2}=k_{3}$, then there are extra isometries.
By composing with an element of $\mathcal{D}_{s}$, we can 
assume that $A_{ij}$ for $i,j=2,3$ is a rotation, but the angle of 
rotation may still depend on the point. By composing
with an element of $\mathcal{D}_{e}$, we can assume that 
$\phi(e)=e$ and $A_{ij}(e)$ is diagonal with $A_{22}(e)=A_{33}(e)=1$. 
If $A_{11}(e)=1$ we are done, so the problem is to exclude the
possibility $A_{11}(e)=-1$. This will be done below.

iii) Assume now that $R_{1}=R_{2}$ or $R_{1}=R_{3}$. If we are lucky,
the relevant $k_{i}$:s are different and we are done, but we
cannot assume that. Since $R_{2}\neq R_{3}$, we do however get the
conclusion that $\phi$ leaves invariant, or at worst changes the sign, 
of one of $\tilde{E}_{2}$ and $\tilde{E}_{3}$. This case will be
pursued further below.

Since $\phi$
preserves the Levi-Civita connection, we get the conclusion that
\[
\nabla_{E_{i}'}E_{j}'=\phi_{*d}\nabla_{\tilde{E}_{i}}\tilde{E}_{j}=
<\nabla_{\tilde{E}_{i}}\tilde{E}_{j},\tilde{E}_{m}>A_{mn}\tilde{E}_{n}.
\]
If we express $E_{i}'$ and $E_{j}'$ in the left hand side in terms
of the basis $\tilde{E}_{k}$, take the scalar product of the result
with $\tilde{E}_{l}$ and finally multiply the result with $A$ in a suitable
fashion, we obtain
\begin{equation}\label{eq:regret}
\tilde{E}_{o}(A_{jl})+A_{jm}<\nabla_{\tilde{E}_{o}}\tilde{E}_{m},
\tilde{E}_{l}>=
<\nabla_{\tilde{E}_{i}}\tilde{E}_{j},\tilde{E}_{m}>A_{io}A_{ml}.
\end{equation}
Consider the case ii). Then $A_{1l}$ is constant since
$R_{1}\neq R_{2}=R_{3}$. Letting $j=1,o=2$ and $l=3$ in (\ref{eq:regret})
we get
\[
A_{11}(e)<\nabla_{\tilde{E}_{2}}\tilde{E}_{1},\tilde{E}_{3}>=
<\nabla_{\tilde{E}_{2}}\tilde{E}_{1},\tilde{E}_{3}>A_{22}(e)A_{33}(e).
\]
Since $A_{22}(e)=A_{33}(e)=1$ and
$<\nabla_{\tilde{E}_{2}}\tilde{E}_{1},
\tilde{E}_{3}>=
-n_{1}/2$, we get the conclusion that $A_{11}(e)=1$.

Assume that $R_{1}=R_{2}$. Let $(l,o,j)=(2,2,3)$ in
(\ref{eq:regret}). Since $A_{32}=0$ by the orthogonality of 
eigenvectors of $\tilde{R}$ with different eigenvalues and since
$<\nabla_{\tilde{E}_{i}}\tilde{E}_{j},\tilde{E}_{m}>=0$ unless
$i,j,m$ are all different, we get
\[
<\nabla_{\tilde{E}_{i}}\tilde{E}_{3},\tilde{E}_{m}>A_{i2}A_{m2}=0.
\]
Thus $(n_{1}-n_{2})A_{12}A_{22}=0$.
Since $n_{1}$ is negative and $n_{2}$ is positive, we conclude
that $A_{12}A_{22}=0$. Since $A$ is an orthogonal matrix we have
$A_{12}=\pm A_{21}$ and $A_{11}=\pm A_{22}$. In any case, the
matrix $A_{ij}$ is a constant matrix, and since the commutator
can be expressed in terms of the Levi-Civita connection,
(\ref{eq:ntrans}) implies 
\begin{equation}\label{eq:wellwell}
\frac{1}{\det A}{}^{t}AnA=n.
\end{equation}
Since $A_{3j}=0$ if $j\neq 3$ and $A\in O(3)$, we get the conclusion
that $A\in SO(3)$. Combining this observation with
(\ref{eq:wellwell}), we exclude 
the possibility $A_{12}\neq 0$. Thus the only possibility
is the situation when $A$ is diagonal with diagonal elements $\pm 1$
and determinant one. The argument concerning the case when
$R_{1}=R_{3}\neq R_{2}$ is similar. $\hfill\Box$



Let us sort out the relation between two canonical bases for a given
set of Bianchi VIII initial data. Let $\mathcal{S}$ be the 
subgroup of $SO(3)$ generated by the three diagonal matrices that have
two minus signs and one plus sign on the diagonal, and the matrix with
$a_{11}=-1$, $a_{23}=1$, $a_{32}=1$ and the remaining components
zero. We say that $\{ e_{i}\} \sim \{ e_{i}'\}$ if there is a matrix
$A\in \mathcal{S}$ (with components $a_{ij}$)
such that $e_{i}=a_{ij}e_{j}'$. Let $\mathcal{S}_{N}$ be the subgroup
generated by $\mathcal{S}$ and the matrices $A$ with components 
$a_{11}=1$, $a_{1i}=a_{i1}=0$ if $i\neq 1$ and $a_{ij}$, $i,j=2,3$,
a rotation matrix. We say that $\{ e_{i}\} \sim_{N} \{ e_{i}'\}$ if 
there is a matrix $A\in \mathcal{S}_{N}$ such that
$e_{i}=a_{ij}e_{j}'$.

\begin{lemma}\label{lemma:sim}
Consider two canonical bases $\{ e_{i}\}$ and $\{ e_{i}'\}$ with
respect to Bianchi VIII initial data $(G,g,k)$. Then they have the
same orientation. If one of the bases is of NUT type, then
$\{ e_{i}\}\sim_{N} \{ e_{i}'\}$. In particular, the other basis is 
of NUT type and $k_{i}=k_{i}'$ and $n_{i}=n_{i}'$, where $k_{i}$ and
$k_{i}'$ are the diagonal components of $k$ with respect to the
different bases and similarly for $n_{i}$ and $n_{i}'$. If one of the
bases is not of NUT type, then $\{ e_{i}\}\sim \{ e_{i}'\}$.
\end{lemma}

\textit{Proof}. 
Assume that $\{ e_{i}\}$ and 
$\{e_{i}'\}$ are two canonical bases corresponding to the same initial
data. Then there is an orthogonal matrix $A$ with components $a_{ij}$
such that 
\[
e_{i}=a_{ij}e_{j}'.
\]
If we take the determinant of (\ref{eq:ntrans}), bearing in mind that 
$n$ and $n'(=\hat{n})$ both have negative determinant, we get the
conclusion that $\det A>0$, whence $A\in SO(3)$. In other words, two
canonical bases have the same orientation. Thus
\begin{equation}\label{eq:ntra}
n'={}^{t}AnA.
\end{equation}
Assume for the sake of argument that we know that $e_{1}=\pm e_{3}'$.
Then (\ref{eq:ntra}) implies 
\[
n_{3}'=\sum_{k}n_{k}a_{k3}^{2}=n_{1},
\]
contradicting the definition of a canonical basis. Similarly, the
assumption that $e_{1}=\pm e_{2}'$ leads to the conclusion that 
$n_{2}'=n_{1}$, and thus to a contradiction. 

Assume that $e_{i}$ is a NUT basis. By (\ref{eq:eigen}), we know
that $R_{1}\neq R_{2}=R_{3}$, with notation as in the proof of
Proposition \ref{prop:isom}. Let us use the notation 
$R_{i}'=\mathrm{Ric}(e_{i}',e_{i}')$, and note that the $e_{i}$ and
the $e_{i}'$ are eigenvectors of $\tilde{R}$ and $\tilde{k}$, defined
in the proof of Proposition \ref{prop:isom}. If $R_{1}'\neq R_{1}$, then
either $R_{2}'=R_{1}$ or $R_{3}'=R_{1}$, i.e. either $e_{1}=\pm
e_{2}'$ or $e_{1}=\pm e_{3}'$. By the above argument, this cannot be. 
Thus $R_{1}'=R_{1}$ and thus $R_{2}'=R_{3}'=R_{2}$ and $e_{1}=\pm
e_{1}'$. We conclude that $A\in \mathcal{S}_{N}$. Note that this
implies $k_{i}=k_{i}'$ and $n_{i}=n_{i}'$ due to (\ref{eq:ntra}) and the
fact that $k_{2}=k_{3}$ and $n_{2}=n_{3}$.

Let  $\{ e_{i}\}$ and $\{ e_{i}'\}$ be canonical bases corresponding 
initial data which is not of NUT type. Assume that the $R_{i}$ are all
different or that the $k_{i}$ are all different. Then the $e_{i}$ are
permutations of the $e_{i}'$, possibly with some sign. If $e_{1}$
does not coincide with $\pm e_{1}'$, we get a contradiction as before.
Since the matrix relating the different bases has to be in $SO(3)$,
we get the conclusion that $\{ e_{i}\} \sim \{e_{i}'\}$. In what
follows, we can thus assume that two of the eigenvalues of $\tilde{R}$
coincide and that two of the eigenvalues of $\tilde{k}$ coincide. 

1) Assume $R_{1}\neq R_{2}=R_{3}$. Then $k_{2}\neq k_{3}$. Assuming
$R_{1}'\neq R_{1}$ implies $e_{1}=\pm e_{2}'$ or $e_{1}=\pm e_{3}'$
and we get a contradiction as before. Thus $e_{1}=\pm e_{1}'$ and 
$k_{1}=k_{1}'$. We may by the above assume that $k_{1}=k_{2}$ or
$k_{1}=k_{3}$. By applying an element of $\mathcal{S}$ to 
$\{ e_{i}'\}$ and one to $\{ e_{i}\}$, we can assume that $k_{3}'\neq
k_{2}'=k_{1}'$ and $k_{1}=k_{2}\neq k_{3}$. Consequently, 
$e_{3}=\pm e_{3}'$. Thus $\{ e_{i}\}\sim \{ e_{i}'\}$.

2) Assume $R_{1}=R_{2}\neq R_{3}$ or $R_{1}=R_{3}\neq R_{2}$. The 
second case can be transformed into the first by applying a matrix
in $\mathcal{S}$ to the basis. If $R_{1}'=R_{3}$, we get 
$e_{1}'=\pm e_{3}$, which leads to a contradiction similarly to 
before. Thus $R_{1}'=R_{1}$, and by applying a matrix in $\mathcal{S}$
to $\{ e_{i}'\}$ we can assume $R_{1}'=R_{2}'\neq R_{3}'$ and that
$e_{3}=e_{3}'$. Consequently, $e_{1}',e_{2}'$ is a rotation of 
$e_{1},e_{2}$. Combining this observation with (\ref{eq:ntra}),
one can conclude that $\{ e_{i}\}\sim \{ e_{i}'\}$.

Since the $R_{i}$ cannot all coincide, the lemma follows. $\hfill\Box$

\begin{lemma}\label{lemma:isoa}
Consider Bianchi VIII initial data $(G,g,k)$ where $G$ is 
simply connected. Let $\{ e_{i}\}$ be a canonical basis. If the
basis is not of NUT type, the isometry group of is generated 
by the left translations and the three Lie group isomorphisms
which on the Lie algebra level change the sign of two of the
$e_{i}$ and leave the remaining one unchanged. If the basis is
of NUT type, the isometry group consists of the above mentioned
diffeomorphims plus the Lie group isomorphims which on the Lie 
algebra level are rotations of $e_{2},e_{3}$.
\end{lemma}

\textit{Proof}. Multiply the $e_{i}$ by positive constants such
that the resulting frame, say $\{ e_{i}'\}$, has 
$n=\mathrm{diag}(-1,1,1)$. Let $(g',k')$ be data on $\tsl$ with
$g(e_{i}',e_{j}')=g'(E_{i},E_{j})$ and 
$k(e_{i}',e_{j}')=k'(E_{i},E_{j})$. Then the map $\phi$ from the 
Lie algebra of $\tsl$ to the Lie algebra of $G$ defined by
$\phi(E_{i})=e_{i}'$ yields a Lie algebra isomorphism. Since $\tsl$ is
a simply connected Lie group, there is an analytic group isomorphism
\[
\psi:\tsl \rightarrow G
\]
such that
\[
\exp\circ\phi=\psi\circ\exp,
\]
see e.g. \cite{lie}. Observe that if $e\in\tsl$
represents the identity and $h$ and arbitrary element, then
\[
e_{i,\psi(h)}'=\frac{d}{dt}\left( \psi(h)\exp[t e_{i,e}']\right)(0)=
\frac{d}{dt}\psi(h\exp[t E_{i,e}])(0)=\psi_{*}E_{i,h}.
\]
Thus
\[
\psi^{*}g(E_{i,h},E_{j,h})=g(e_{i,\psi(h)}',e_{j,\psi(h)}')
=g'(E_{i,h},E_{j,h})
\]
and similarly for $k$. Thus $\psi^{*}g=g'$ and $\psi^{*}k=k'$.
The lemma now follows from Proposition \ref{prop:isom}. $\hfill\Box$

\begin{lemma}\label{lemma:sia}
Consider Bianchi VIII initial data $(G,g,k)$ and let $\{ e_{i}\}$
and $\{ e_{i}'\}$ be two canonical bases. Then the class A
developments associated with these canonical bases coincide.
Assume now that $G$ is simply connected. 
Then there are initial data $(\tsl,g',k')$ such that $g'$ and $k'$
are diagonal with respect to $E_{i}$ and a Lie group isomorphim
$\psi:\tsl\rightarrow G$ such that $\psi^{*}g=g'$ and 
$\psi^{*}k=k'$. Furthermore the intervals of existence appearing
in the corresponding class A developments are the same and $(1,\psi)$
is an isometry of these developments.
\end{lemma}

\textit{Proof}. Let us first consider two canonical bases $\{ e_{i}\}$
and $\{ e_{i}'\}$ corresponding to the same initial data $(G,g,k)$,
where $G$ is of Bianchi type VIII but is not necessarily simply
connected. We wish to prove that the corresponding class A developments 
coincide. The construction of the class A development, given a 
canonical basis $\{ e_{i}\}$ proceeds as follows (see pp. 3798--99 of
\cite{jag3}). We define $\theta(0)=-\mathrm{tr}_{g}k$ and
$\sigma_{i}(0)=-k_{i}+\theta/3$ and call the diagonal components of
the $n$ matrix of the basis $n_{i}(0)$. We call the collection
$[n_{i}(0),\sigma_{i}(0),\theta(0)]$ Ellis and MacCallum initial data
corresponding to the basis $\{ e_{i}\}$. Then we solve (11), (13) and
(14) of \cite{jag3} to obtain an existence interval $I$ on which
$n_{i}$, $\sigma_{i}$ and $\theta$ are defined. The $a_{i}$ in
(\ref{eq:metric}) are given by
\begin{equation}\label{eq:aia}
a_{i}(t)=\exp[\int_{0}^{t}(\sigma_{i}+\frac{1}{3}\theta)ds]
\end{equation}
and the $\xi^{i}$ in (\ref{eq:metric}) are the duals of the $e_{i}$.
Let $A\in SO(3)$, with components $a_{ij}$ be such that
$e_{i}=a_{ij}e_{j}'$. If $A$ has two minus signs on the diagonal and
one plus sign, then the Ellis and MacCallum initial data corresponding
to the different bases coincide. Thus the existence intervals $I$ and
$I'$ coincide and the $\theta$ and $\sigma_{i}$ coincide. By
(\ref{eq:aia}), we see
that the $a_{i}$ coincide. Thus the corresponding developments
coincide. Assume that $A$ has $a_{11}=-1$, $a_{23}=1$, $a_{32}=1$ and
that the remaining components are zero. On the level of the Ellis and
MacCallum initial data, this corresponds to interchanging $n_{2}(0)$
with $n_{3}(0)$ and similarly for the $\sigma_{i}(0)$. Considering
(11), (13) and (14) of \cite{jag3} one can see that this operation
can be extended to the entire solution. That is, interchanging $n_{2}$
$n_{3}$ and $\sigma_{2}$ $\sigma_{3}$ at the same time maps solutions
to solutions. Consequently $I=I'$, and since we get $a_{2}=a_{3}'$ and
$a_{3}=a_{2}'$, we conclude that the developments coincide (note that 
$e_{2}=e_{3}'$ and that $e_{3}=e_{2}'$). Finally, assume that one of
the bases is a NUT basis and that $a_{11}=1$, $a_{i1}=a_{1i}=0$ if 
$i\neq 1$ and that $a_{ij}$, $i,j=2,3$ is a rotation matrix. 
Then the Ellis and MacCallum initial data have to coincide, so that 
$I=I'$. Considering (11), (13) and (14) of \cite{jag3}, one concludes
that $\sigma_{2}=\sigma_{3}$ so that $a_{2}=a_{3}=a_{2}'=a_{3}'$ for
all time. It is then easy to see that the developments coincide. 

Assuming that $G$ is simply connected, the proof that one can find
a Lie group isomorphism $\psi:\tsl\rightarrow G$ and initial data
$(\tsl,g',k')$ as in the statement of the lemma is similar to the
arguments presented in the proof of Lemma \ref{lemma:isoa}. The map
$\psi$ will on the Lie algebra level map one canonical basis to
another. Therefore, the corresponding Ellis and MacCallum initial data
will coincide. Consequently the different intervals of existence 
and $a_{i}$ coincide. Thus, $(1,\psi)$ defines an isometry of
developments. $\hfill\Box$

\section{Compactifications}

The essence of the arguments in this section are taken from
\cite{thurston}, Proposition 4.7.2 and Corollary 4.7.3 and from
\cite{bredon}, Theorem 12.8. 

\begin{lemma} 
Assume that $\Gamma$ is a 
free and properly discontinuous subgroup of $\mathcal{D}_{e}$ such 
that $\tsl/\Gamma$ is compact.
Then $p(\Gamma)$ is a discrete subgroup of the isometry group of
hyperbolic space.
\end{lemma}
\textit{Remark}. We here assume that $\mathrm{Isom}(H^{2})$ has been
given a Lie group structure by identifying the orientation preserving
part with $\psl$. When we say that $p(\Gamma)$ is discrete, we mean
that it is a discrete subset of the topological space
$\mathrm{Isom}(H^{2})$. 

\textit{Proof}. Note that we have the short exact sequence
(\ref{eq:exact}). Let $U$ be an open connected neighbourhood of
the identity of $\mathcal{D}_{e}$ such
that $[U,U]\subseteq V$, where $V\cap \Gamma=\{ e\}$. Here
\[
[U,U]=\{ ghg^{-1}h^{-1}:g, h\in U\}.
\]
Two elements of $g,h\in\Gamma$ which project to $p(U)$ commute.
The reason is that there are elements $t,s\in\mathbb{R}$ such
that $tg,sh\in U$. Since the elements of $\mathbb{R}$ are central
in $G$, we have
\[
(tg)(sh)(tg)^{-1}(sh)^{-1}=ghg^{-1}h^{-1}\in [U,U]\cap \Gamma
\subseteq \{ e\}.
\]
The subgroup $H$ of $\mathrm{Isom}(H^{2})$ generated by $p(\Gamma)\cap
p(U)$ is thus abelian. Furthermore, if $\gamma\in \Gamma$, choose
an open connected neighbourhood $U_{\gamma}\subseteq U$ of the
identity such that 
$[\gamma,U_{\gamma}]\subseteq V$. Then the group $H_{\gamma}$
generated by $p(\Gamma)\cap p(U_{\gamma})$ satisfies $H_{\gamma}
\subseteq H$, so that it is abelian, and every element of 
$H_{\gamma}$ commutes with $p(\gamma)$. Note that since $U$ is a 
connected neighbourhood of the identity,
$H$ must consist of orientation preserving isometries
of hyperbolic space. In other words $H_{\gamma}\subseteq H
\subseteq\psl$. We now wish to prove that 
if $p(\Gamma)$ is not discrete, then $\bar{H}$ and $\bar{H}_{\gamma}$,
where the closure is taken in $\psl$,
are equal and constitute a one parameter subgroup of $\psl$.
Consider $\sl$. We have the exponential map from the vector space of
tracefree $2\times 2$ matrices $T_{2}$ to $\sl$. Locally around the origin, 
this is a diffeomorphism. The covering homomorphism $\pi:\sl\rightarrow\psl$
is also locally a diffeomorphism. Composing, we get the local 
diffeomorphism $\pi\circ \exp$ from a neighbourhood of the origin
of $T_{2}$ to a neighbourhood of the identity of $\psl$. Note that
the inverse image of $\bar{H}$ under $\pi\circ\exp$ is a closed
subset $S$ of $T_{2}$ with the property that if $v\in S$, then
$nv\in S$ for all $n\in\mathbb{Z}$. Since $p(U)$ is an open
neighbourhood of the identity, the assumption that $p(\Gamma)$
is not discrete leads to the conclusion that the origin in $T_{2}$ is
not isolated in $S$. Note that we can give $T_{2}$ a norm $\|\cdot\|$ 
through the isomorphism of $T_{2}$ with $\mathbb{R}^{3}$. For 
$n\in\mathbb{N}$, let $x_{n}\in B_{1/n}(0)\cap S-\{ 0\}$ and let 
$k_{n}\geq n$ be an integer such that $1\leq \|k_{n}x_{n}\|\leq 2$.
We can assume that $y_{n}=k_{n}x_{n}$ converges to an element $y$.
Let $t\neq 0$ be a real number. There is a sequence of integers
$r_{n}$ such that $|r_{n}/k_{n}-t|\leq 1/n$. Since 
$r_{n}x_{n}\rightarrow ty$, we get the conclusion that $ty\in S$.
Thus $S$ contains the line $\mathbb{R}y$, and $\bar{H}$ contains
a one parameter subgroup of $\psl$. The same argument yields the 
conclusion that $\bar{H}_{\gamma}$ contains a one parameter subgroup.
Since $\bar{H}$ is abelian, orientation preserving and contains a
non-trivial orientation preserving isometry we can apply Lemma
\ref{lemma:asg}. The only possibilities for $\bar{H}$ are, up to
conjugation, rotations around the origin in the disc model, the group
of isometries of the form (\ref{eq:hyp}) and the translations by real
numbers in the upper half plane. Since $\bar{H}_{\gamma}\subseteq 
\bar{H}$ also contains a one parameter subgroup of isometries, the 
two groups must coincide. Since $p(\gamma)$ commutes with 
$\bar{H}_{\gamma}=\bar{H}$ for all $\gamma\in\Gamma$, we get the
conclusion that $p(\Gamma)$ commutes with $\bar{H}$. Consequently,
$p(\Gamma)$ is a subgroup of one of the groups mentioned in Lemma
\ref{lemma:asg}.

Assume that $\tsl/\Gamma$ is compact and let $g$ be a metric on 
$\tsl$ whose isometry group contains $\mathcal{D}_{e}$. 
Then there is an $r>0$ such that for any $p\in\tsl$ the projection
$\pi : \tsl\rightarrow \tsl/\Gamma$ is injective when restricted to
$B_{r}(p)$, where the ball is defined using the topological metric
induced by $g$. The reason
is that if this were not the case, then there would be $p_{n}\in\tsl$
and $0<r_{n}\leq 1/n$ such that $\pi$ would not be injective on
$B_{r_{n}}(p_{n})$. Since the quotient is compact, we can assume that 
$\pi(p_{n})$ converges to a point $q$. Let $p\in\tsl$ satisfy
$\pi(p)=q$ and let $r>0$ be such that $\pi$ is injective on
$B_{r}(p)$. This means that $\pi$ is injective on 
$\gamma B_{r}(p)=B_{r}(\gamma p)$ for every $\gamma\in \Gamma$ and
that $\gamma B_{r}(p)\cap B_{r}(p)\neq\emptyset$ implies 
$\gamma=e$. Since $\pi[B_{r}(p)]$ contains an open neighbourhood of 
$q$, we get the conclusion that $\pi[B_{r_{n}}(p_{n})]\subseteq
\pi[B_{r}(p)]$ for $n$ large enough. Thus $B_{r_{n}}(p_{n})
\subseteq \gamma B_{r}(p)$ for some $\gamma\in \Gamma$. We have 
a contradiction. 

Let us define a map $\tilde{p}:\tsl\rightarrow D$ by
\begin{equation}\label{eq:ptdef}
\tilde{p}(h)=p[L_{h}](0).
\end{equation}
If $\psi\in\mathcal{D}_{e}$, we wish to prove that 
\begin{equation}\label{eq:tpprel}
\tilde{p}[\psi(h)]=p[\psi]\tilde{p}(h).
\end{equation}
If $\psi=L_{g}$, then (\ref{eq:tpprel}) holds. If $\psi$ is a Lie 
group isomorphism, then
\[
\tilde{p}[\psi(h)]=p[L_{\psi(h)}](0)=p[\psi L_{h}\psi^{-1}](0)=
p[\psi]p[L_{h}]p[\psi^{-1}](0)=p[\psi]\tilde{p}(h),
\]
since $p[\psi^{-1}](0)=0$ in this case. Since $p$ is a homomorphism
and the elements of $\mathcal{D}_{e}$ can be written as combinations
of left translations and Lie group isomorphisms, we get the conclusion
that (\ref{eq:tpprel}) holds in general. 

As mentioned above, there are three possibilities for $p(\Gamma)$.
Assume there is an isometry $\chi$ such that $\chi\ p(\Gamma)\chi^{-1}$
is a subgroup of the rotations around the origin in the disc model.
The presence of $\chi$ is only a technical nuisance, and for that 
reason we will assume it to be the identity below.
Let $\gamma_{t}\in\psl$ be defined by (\ref{eq:hyp}) and let 
$\tga_{t}\in\tsl$ satisfy $p(L_{\tga_{t}})=\gamma_{t}$. Let
$r>0$ be such that $\pi$ is injective on $B_{r}(h)$ $\forall
h\in\tsl$. Assume we have found $t_{1},...,t_{n}$ such that $\pi$ is
injective on $\cup_{k=1}^{n}\tga_{t_{k}} B_{r}(e)$. We wish to find a
$t_{n+1}$ such that we can increase the union to $n+1$ and preserve 
the injectivity. Since $\pi$ is injective on $\tga_{t_{n+1}}B_{r}(e)$
by the choice of $r$, the only conceivable problem would be the
existence of a $\gamma\in\Gamma$ and a $k\leq n$ such that 
$\tga_{t_{n+1}}B_{r}(e)\cap\gamma\tga_{t_{k}}B_{r}(e)\neq \emptyset$. 
However, this would imply
\[
p[L_{\tilde{\gamma}_{t_{n+1}}}]\tilde{p}[B_{r}(e)]\cap 
p[\gamma]p[L_{\tilde{\gamma}_{t_{k}}}]
\tilde{p}[B_{r}(e)]\neq\emptyset.
\]
Note however that 
$\cup_{k=1}^{n}p[L_{\tilde{\gamma}_{t_{k}}}]\tilde{p}[B_{r}(e)]$
is inside a ball of radius strictly less than $1$ in $D$ and that 
$p[\gamma]$
leaves this ball invariant for $\gamma\in \Gamma$. Choosing
$t_{n+1}$ to be close enough to $1$, one can show that 
$p[L_{\tilde{\gamma}_{t_{n+1}}}]\tilde{p}[B_{r}(e)]$ is outside 
of this ball. Thus
for arbitrary $n\in\mathbb{N}$, one can choose a sequence 
$t_{1},...,t_{n}$ such that $\pi$ is injective on the union of
$\tga_{t_{k}}B_{r}(e)$. Since these sets all have the same fixed
volume, we get the conclusion that the quotient has infinite volume.

The arguments in the other two cases are similar. If $\chi\
p(\Gamma)\chi^{-1}$ is a subgroup of the group of translations by
real numbers in the upper half plane, the isometries that should
replace the $\gamma_{t}$:s in the above argument are the dilations
of the upper half plane. Consider the case when $\chi\
p(\Gamma)\chi^{-1}$ is a subgroup of the isometries of the form
(\ref{eq:hyp}). The $\gamma_{t}$ should in this case be replaced with
$\psi_{t}$ defined by
\[
\psi_{t}(z)=\frac{z+it}{1-itz}.
\]
The reason is the following. Let $U$ be contained in a compact subset
of $D$. Then $\psi_{t}(U)$ tends to the point $i$ as $t\rightarrow 1$.
On the other hand, if $\phi_{t}$ is defined by (\ref{eq:hyp}), then
the imaginary part of $\phi_{t}(U)$ can be bounded from above by a 
number strictly less than $1$ independent of $t$. $\hfill\Box$

\begin{lemma}\label{lemma:comp}
Assume that $\Gamma$ is a free and properly discontinuous
subgroup of $\mathcal{D}_{e}$ such that $\tsl/\Gamma$ is compact.
As noted above, $p(\Gamma)$ is a discrete subgroup of the isometry
group of hyperbolic space. Furthermore, $D/p(\Gamma)$ is compact
and $\Gamma\cap \mathbb{R}$ is an infinite cyclic group. Here,
$\mathbb{R}$ is to be interpreted as the subgroup of translations.
Finally, $\tsl/\Gamma$ is a Seifert fibred space with a compact hyperbolic
orbifold as a base. 
\end{lemma}

\textit{Proof}.
Let us prove that if $z\in D$, then $p(\Gamma)z$ is a closed discrete
set. It is enough to prove that for every $\zeta\in D$, there is an
open set $U\subseteq D$ such that $\zeta\in U$ and $U$ only contains
a finite number of elements in $p(\Gamma)z$. Assume that 
$p(\gamma_{n})z\rightarrow \zeta$ and that all the $p(\gamma_{n})z$
are distinct. We can choose a subsequence so that $p(\gamma_{n})$
are all orientation preserving or orientation reversing. If 
they are all orientation reversing, we can furthermore consider the 
sequence $p(\gamma_{1}\gamma_{n})$ instead. This sequence will have
the same properties, with $\zeta$ replaced by $p(\gamma_{1})\zeta$, 
so we can assume that the original sequence 
consists of orientation preserving isometries.
Note that we can identify $\psl$ with $UD$ and that the 
above assumption implies that if one projects $p(\gamma_{n})$ to $D$,
one remains in the same compact set for all
$n$. Thus, $p(\gamma_{n})$ belongs to a compact subset of $\psl$.
Thus we can assume that the sequence converges. Consequently,
$p(\gamma_{n}\gamma_{n+1}^{-1})$ is a sequence of elements of 
$p(\Gamma)$ separate from the identity, converging to the identity. 
This is not possible since $p(\Gamma)$ is discrete. 

In what follows it will be useful to know that if $p[\phi]$ has
a fixed point and is orientation reversing, then $\phi$ has a 
fixed point. Assume to that end that $p[\phi](\zeta)=\zeta$ and that
$p[\phi]$ is orientation reversing. Let $\chi$ be an element of the 
orientation preserving isometry group of hyperbolic space such that 
$\chi^{-1}(0)=\zeta$ and let $\hat{\chi}$ satisfy $p(\hat{\chi})=\chi$. 
If we let $\hat{\phi}=\hat{\chi}\phi\hat{\chi}^{-1}$, then 
$p[\hat{\phi}](0)=0$. Due to the representation
(\ref{eq:representation}) we have 
\[
\hat{\phi}=\phi_{2} T_{\theta}L_{(\alpha,z)}
\]
with notation as in the proof of Lemma \ref{lemma:isom}. 
Thus $z=0$ and we conclude that $(-(\alpha+\theta)/2,0)$ is a 
fixed point of $\hat{\phi}$. Consequently, $\phi$ has a fixed point. 

Consider the subgroup $p(\Gamma)_{o}$ of $p(\Gamma)$ fixing the origin
$o\in D$. Since $\Gamma$ is free and properly discontinuous, we
conclude by the above that $p(\Gamma)_{o}$ consists of rotations
around the origin. Since it is finite due to the 
discreteness of $p(\Gamma)$, it is generated by a rotation
through an angle $2\pi q$ where $q\in\mathbb{Q}$. Let 
\[
l=\inf\{ d(o,p(\gamma) o)|p(\gamma) o\neq o,\ \gamma\in\Gamma\}.
\]
Since $p(\Gamma)o$ is a discrete set, $l>0$. Let $0<r<l/2$. Then
if $p(\gamma)B_{r}(o)\cap B_{r}(o)\neq\emptyset$, $p(\gamma)$ has
to belong to $p(\Gamma)_{o}$. Let $U\subseteq B_{r}(o)$ be
an open and non-empty set contained in a sector of an angle less that 
$2\pi q$.

Consider (\ref{eq:exact}). The subgroup $K$ of $\mathbb{R}$ obtained
by taking the inverse image of $\Gamma$ under the injection from
$\mathbb{R}$ to $\mathcal{D}_{e}$ has to be discrete due to the
discreteness of 
$\Gamma$. Thus it is trivial or an infinite cyclic group. Assume
that it is trivial. Then $p$ restricted to $\Gamma$ is 
an isomorphism. Let $\tilde{U}=\mathbb{R}\times U$. Assume
$\phi \tilde{U}\cap\tilde{U}\neq\emptyset$ for
$\phi\in\Gamma$. Then $\phi(\alpha_{1},z_{1})=(\alpha_{2},z_{2})$ for
$z_{i}\in U$, $i=1,2$. However,
\[
\tilde{p}(\alpha_{2},z_{2})=p[L_{(\alpha_{2},z_{2})}](0)=
\pi_{2}[L_{(\alpha_{2},z_{2})}(0,0)]=z_{2},
\]
where $\tilde{p}$ is defined in (\ref{eq:ptdef}) and, similarly,
\[
\tilde{p}[\phi(\alpha_{1},z_{1})]=p[\phi](z_{1}).
\]
Thus $p[\phi]U\cap U\neq\emptyset$. Consequently,
$p[\phi]=\mathrm{Id}$ and thus $\phi=\mathrm{Id}$, since $p$
restricted to $\Gamma$ is
injective. Take any ball $B$ in $\tilde{U}$. 
By translations by $\mathbb{R}$, one can then
construct an infinite number of disjoint balls in $\tilde{U}$ all
of which are isometric to $B$. This implies that the volume of 
$\tsl/\Gamma$ is infinite. Consequently, $K$ has to be an infinite
cyclic group. Let $\pi_{O}:D\rightarrow D/p(\Gamma)$. Then 
$\pi_{O}\tilde{p}[\phi(g)]=\pi_{O}\tilde{p}(g)$ for $\phi\in \Gamma$. 
Thus we get a continuous map from $\tsl/\Gamma$ onto $D/p(\Gamma)$. In
other words, $D/p(\Gamma)$ is compact. 

Let us prove that the quotient is Seifert fibred. Recall that $\tsl$
is $\mathbb{R}\times D$ topologically and denote the generator of 
$\Gamma\cap \mathbb{R}$ by $\delta>0$. Let $(\a,z)\in \mathbb{R}\times
D$ and assume that only the identity element of $p(\Gamma)$ fixes
$z$. Let $B_{\e}(z)\subseteq D$ be an open ball such that 
$p(\gamma)B_{\e}(z)\cap B_{\e}(z)\neq\emptyset$ for $\gamma\in \Gamma$
implies $p(\gamma)=\mathrm{Id}$. Consider the set 
$U=[\a,\a+\delta]\times B_{\e}(z)$. Assume that $\gamma p_{1}=p_{2}$
for $p_{i}\in U$ and $\gamma\in\Gamma$. Then
$p(\gamma)\tilde{p}(p_{1})=\tilde{p}(p_{2})$. Since $\tilde{p}(p_{i})$
are both in $B_{\e}(z)$ we get the conclusion that $p(\gamma)=
\mathrm{Id}$. Consequently $\gamma$ is a translation by $\delta$. 
In other words, the image of $U$ in $\tsl/\Gamma$ is a cylinder. 
Thus there is a neighbourhood of the image of $(\a,z)$ homeomorphic
to a cylinder. Assume that the subgroup $p(\Gamma)_{z}$ of 
$p(\Gamma)$ fixing $z$ is non-trivial. Let $\chi$ be an orientation
preserving isometry of hyperbolic space such that $\chi^{-1}(0)=z$
and let $\hat{\chi}$ be such that $p(\hat{\chi})=\chi$. Assume that 
$p(\gamma)\in p(\Gamma)_{z}$. Then $\chi p(\gamma)\chi^{-1}(0)=0$.
Since $p(\gamma)$ is orientation preserving, we conclude that 
\[
\hat{\chi}\gamma\hat{\chi}^{-1}=T_{\theta}L_{(\beta,0)}
\]
using (\ref{eq:representation}). Let $\delta$ be defined by
\[
\delta=\inf \{ \theta+\beta: \hat{\chi}^{-1}T_{\theta}L_{(\beta,0)}
\hat{\chi}\in \Gamma, \theta+\beta>0\}.
\]
Since $\Gamma$ is free and properly discontinuous, $\delta>0$ and
there are $\theta$ and $\beta$ such that $\delta=\theta+\beta$
and $\hat{\chi}^{-1}T_{\theta}L_{(\beta,0)}\hat{\chi}\in \Gamma$. 
Note that $\theta$ and $\beta$ are unique, since if $\theta_{1}$
and $\beta_{1}$ would have the same properties as $\theta$ and 
$\beta$ but be different from them, then one could construct a map
in $\Gamma$ with a fixed point which would be different from the 
identity. Since $p[T_{\theta}L_{(\beta,0)}]$ is a rotation 
through an angle $\beta$ and $p(\Gamma)$ is discrete, we get the 
conclusion that $\beta=2\pi q$ where $q\in\mathbb{Q}$. Let $\e>0$
be such that if $p(\gamma)B_{\e}(z)\cap B_{\e}(z)\neq\emptyset$
then $p(\gamma)z=z$. Let $U=\hat{\chi}^{-1}\{ [\a,\a+\delta]\times 
B_{\eta}(0)\}$, where $\eta>0$ is such that $\tilde{p}(U)\subseteq 
B_{\e}(z)$. Assume that $\gamma p_{1}=p_{2}$ where 
$p_{i}=\hat{\chi}^{-1}(\a_{i},z_{i})\in U$ and 
$\gamma\in\Gamma$. Then $p(\gamma)\tilde{p}(p_{1})=\tilde{p}(p_{2})$
so that $p(\gamma)z=z$ since $\tilde{p}(p_{i})\in B_{\e}(z)$. Thus
$\gamma=\hat{\chi}^{-1}T_{\theta'}L_{(\beta',0)}\hat{\chi}$, whence
\[
(\a_{1}+\theta'+\beta',e^{i\beta'}z_{1})=(\a_{2},z_{2}).
\]
Since $(\a_{i},z_{i})\in [\a,\a+\delta]\times B_{\eta}(0)$, we get the
conclusion that $(\theta',\beta')=\pm (\theta,\beta)$ or 
$(\theta',\beta')=(0,0)$. Consequently, the image of $U$ in
$\tsl/\Gamma$ is homeomorphic to a cylinder on which the ends have
been identified after a rotation by
an angle $2\pi q$. We conclude that $\tsl/\Gamma$ is a Seifert fibred
space. $\hfill\Box$

\textit{Proof of Theorem \ref{thm:seifert}}. 
Let $(G,g,k)$ be Bianchi VIII initial data with $G$ simply connected
and assume that $\Gamma$ is a free and properly discontinuous subgroup 
of the isometry group of the initial data such that $G/\Gamma$ is
compact. Let $g'$, $k'$ and $\psi$ be as in the statement of Lemma
\ref{lemma:sia}. Then $\Gamma'=\psi^{-1}\Gamma\psi$ is a free and properly 
discontinuous group of isometries of $(\tsl,g',k')$, so that 
$\Gamma'\subset \mathcal{D}_{e}$. As was noted in the introduction,
$(1,\Gamma)$ is a free and properly discontinuous group of isometries
of the class A development due to Lemma \ref{lemma:isoa} and similarly
for $\Gamma'$. The isometry $(1,\psi)$ between the Bianchi class A
developments induces an isometry $(1,\psi')$ between the compactified
class A developments $I\times \tsl/\Gamma'$ and $I\times G/\Gamma$. 
Since $\Gamma'\subseteq \mathcal{D}_{e}$, Lemma \ref{lemma:comp}
implies that $\tsl/\Gamma'$ is Seifert fibred. Furthermore, when one 
unwraps the circle fibres, they correspond to $E_{1}$. Due to the 
existence of the isometry $(1,\psi')$ we get the conclusion that 
$G/\Gamma$ is Seifert fibred. Since $\psi$ takes
$E_{1}$ to $e_{1}$, where $e_{1}$ is the first element of a canonical
basis for $(G,g,k)$, see the proofs of Lemma \ref{lemma:isoa} and 
\ref{lemma:sia}, we get the conclusion that the integral curves of 
$e_{1}$ become the circle fibres in the Seifert fibred space under
the compactification. $\hfill\Box$

\end{document}